\documentclass[aps,twocolumn,showkeys]{revtex4-2}
\usepackage{graphicx}
\usepackage{subfigure}
\usepackage{color}
\usepackage{amsmath,amssymb}
\usepackage{xcolor}
\usepackage[colorlinks=true,allcolors=blue]{hyperref}
\usepackage{bbold}
\usepackage[T1]{fontenc}
\usepackage{young}
\usepackage{youngtab}
\usepackage[boxsize=1.07em]{ytableau}
\usepackage{etoolbox}
\usepackage{physics}
\usepackage{braket}
\usepackage{mathtools}

\makeatletter
\newcommand{\pushright}[1]{\ifmeasuring@#1\else\omit\hfill$\displaystyle#1$\fi\ignorespaces}
\newcommand{\pushleft}[1]{\ifmeasuring@#1\else\omit$\displaystyle#1$\hfill\fi\ignorespaces}
\makeatother

\newcommand\mydots{\hbox to 0.9em{.\hss.\hss.}}

\newtheorem{thm}{Theorem}[section]

\newtheorem{prop}[thm]{Proposition}

\def\nn{\nonumber}
\def\ic{\mathrm{i}}

\def\bc{\begin{center}}
\def\ec{\end{center}}
\def\bi{\begin{itemize}}
\def\ei{\end{itemize}}
\def\ba{\begin{array}}
\def\ea{\end{array}}

\newcommand{\la}{\langle}
\newcommand{\ra}{\rangle}

\def\bc {\bar{\beta}}

\def\rmu{\mathrm{U}}
\def\rmsu{\mathrm{SU}}

\def\flux {L}

\begin{document}

\title{Lieb-Mattis ordering theorem of electronic energy levels in the thermodynamic limit}

\author{Manuel Calixto}
\email{calixto@ugr.es}
\affiliation{Department of Applied Mathematics, University of  Granada,
	Fuentenueva s/n, 18071 Granada, Spain}
\affiliation{Institute Carlos I of Theoretical and Computational Physics (iC1), University of  Granada,
	Fuentenueva s/n, 18071 Granada, Spain}

\author{Alberto Mayorgas}
\email{alberto.mayorgasreyes@ceu.es}
\affiliation{Higher Polytechnic School, University CEU Fernando III, Glorieta Cardenal Herrera Oria, 41930 Sevilla, Spain}

\author{Julio Guerrero}
\email{jguerrer@ujaen.es}
\affiliation{Department of Mathematics, University of Jaen, Campus Las Lagunillas s/n, 23071 Jaen, Spain}
\affiliation{Institute Carlos I of Theoretical and Computational Physics (iC1), University of  Granada,
	Fuentenueva s/n, 18071 Granada, Spain}

\date{\today}

\begin{abstract}
	\vspace{1cm}
	\section*{Abstract}
   
   Lieb-Mattis theorem orders the lowest-energy states of total spin $s$ of a system of $P$ interacting fermions. We generalize these predictions to fermionic mixtures of $P$ particles with more than $N=2$  spinor components/species in the thermodynamic limit $P\to\infty$. The lowest-energy state inside each permutation symmetry sector $h$, arising in the $P$-fold tensor product decomposition, is well approximated by a U$(N)$ coherent (quasi-classical, variational) state, specially in the limit $P\to\infty$. In particular, the ground state of the system belongs the most symmetric (dominant Young tableau $h_0$) configuration. We exemplify our construction with the $N=3$ level Lipkin-Meshkov-Glick model, with a previous motivation on pairing correlations and U$(N)$-invariant quantum Hall ferromagnets. In the limit $P\to\infty$, each lowest-energy state within each permutation symmetry sector $h$ undergoes a quantum phase transition for a critical value $\lambda_c(h)$ of the exchange  coupling constant $\lambda$, depending on $h$. This generalizes standard quantum phase transitions and their phase diagrams corresponding to the ground state belonging to the most symmetric sector $h_0$. 

\end{abstract}


\maketitle

\section{Introduction}

Particle physics provides numerous models of interacting multicomponent fermions. The paradigmatic case of $N=2$ spinor components for spin-1/2 fermions extends to $N=3$ color species or $N=6$ flavors when we talk about quarks and leptons. Isospin components also arise in nuclear physics. Here we are interested in the condensed matter and mesoscopic quantum  physics context. In particular, recent advances in $\rmsu(N)$ fermions \cite{Decamp16,PhysRevResearch.2.023059} show how  ultracold atomic  gases can accommodate unitary $\rmsu(N)$ symmetries with larger and larger values of $N$  (see e.g.  \cite{pethick_smith_2008,Lewenstein_2012} for text books and   \cite{Cazalilla_2014,Sowinski19} for  reviews).   Experimental observation of exotic $\rmsu(N)$ physics started with  \cite{doi:10.1126/science.1254978}. 
For instance, the $\rmsu(N)$ generalization of the Hubbard model is realized by fermionic alkaline-earth atomic gases trapped in optical lattices \cite{PhysRevLett.92.170403,PhysRevA.99.063605,Cazalilla_2009,PhysRevA.96.032701}.  In view of recent advancements in cooling, confining, and controlling quantum systems of this complexity, Feynman's initial concepts regarding the simulation of quantum systems and the processing of quantum information are becoming more feasible.

 In certain situations, electrons become multicomponent, as they gain additional ``pseudospin" internal components, apart from the typical $N=2$ spin up ($\uparrow$) and down ($\downarrow$) components. These additional components can be  associated with Dirac valley ($\pm K$) and sublattice ($A, B$) configurations, similar to those found in graphene and other 2D Dirac materials. Multilayer arrangements also introduce additional spinor components to the electron and enhance physical complexity. For example, twisted bilayer  graphene  exhibits interesting superconducting \cite{Bistritzer12233,Cao2018} and entanglement \cite{TBGPhysicaE} properties for some ``magic'' twist angles.  Within the lowest Landau level, one Landau site of a  bilayer quantum Hall system  can accommodate up to $N=4$ electron spinor components/species: $i=(t\uparrow), (t\downarrow), (b\uparrow), (b\downarrow)$, corresponding to the tensor product of spin $\uparrow,\downarrow$ and pseudospin $t,b$ (top and bottom layer), thus resulting in an underlying $\rmu(4)$ symmetry (see e.g. \cite{HamEzawa,Schliemann,PhysRevB.95.235302} for a study of the ground-state structure for filling factor $\nu=2$).  The $\ell$-layer scenario therefore comes with a $\rmu(2\ell)$ symmetry.

In this article, we shall place our discussion of interacting multicomponent  fermion mixtures  in the context of $\rmu(N)$ Quantum Hall  Ferromagnets (QHFs). They generalize the quantum Heisenberg model of interacting spins on a lattice. 
In the low-energy (long wavelength) regime, the Hamiltonian can be expressed in terms of collective $\rmu(N)$-spin operators $S_{ij}, i,j=1,\dots,N$. Semiclassical spin-wave coherent excitations   (usually named ``skyrmions''\cite{Seki-Mochizuki_Skyrmion_2016,JungHoonHan_Skyrmion_2017,Finocchio-Panagopoulos_Skyrmion_2021,Zang_Thesis_Skyrmion_2018}) are traditionally described by classical nonlinear sigma models \cite{AffleckNPB257,AffleckNPB265,AffleckNPB305,Sachdev,Sachdev2,Arovas,CALIXTO1},  generalizing the $\rmsu(2)$   continuum dynamics of  Heisenberg (anti)ferromagnet spin chains \cite{HaldanePLA93,HaldanePLA93-2,HaldanePLA93-3}. For this study, $\rmu(N)$ coherent states\cite{Perelomov,Gazeaubook} are fundamental to replace $\rmu(N)$-spin operators $S_{ij}$ by their coherent state (CS) expectation values $s_{ij}=\la S_{ij}\ra$. Usually, this study is restricted to the fully symmetric sector, where the ground state lies, disregarding other mixed permutation symmetry sectors. For $P$ identical $\rmu(N)$ particles, let us say $P$ symmetric  ``multi-quNits'' \cite{QIP-2021-Entanglement}, this reduces the Hilbert space dimensionality from the tensor product dimension $N^P$ to $\tbinom{P+N-1}{P}$. 

Here we shall address, not only the most symmetric permutation symmetry sector where the ground state system belongs, but all the  irreducible representations (IRs) $h$ of the unitary group $\rmu(N)$ arising in the Clebsch-Gordan direct sum decomposition of the $P$-fold  tensor product representation. The permutation symmetry $S_P$ group of $P$ particles and their related 
Young diagrams turn out to be a useful graphical method to represent these kind of decompositions into different symmetry sectors  
and we shall make extensive use of them in the next Sections. We address the reader to standard  \cite{Thrall1954,Barut,HallBook,CVI} and  pedagogical  \cite{ZHAO} references for the Young diagrammatic method and its relation to the study of $\rmu(N)$ IRs $h$.

Lieb-Mattis theorem provides an ordering of the lower-energy electronic levels inside each permutation symmetry sector (labelled by the total spin $s$), for a general class of symmetric Hamiltonians for $N=2$  fermion (spin) components. 
We intend to generalize these predictions to fermionic mixtures of $P$ particles with more than $N=2$  spinor components/species. Particular examples are found in the recent literature like  \cite{Decamp16,PhysRevResearch.2.023059}   addressing the case of $N\leq 6$-component Fermi gases in tight waveguides. We also want to extend our study to the thermodynamic limit $P\to\infty$, where the phenomenon of  quantum phase transitions (QPTs) arises. For this purpose, we shall make extensive use of U$(N)$ coherent states (see \cite{Perelomov,Gazeaubook,RevModPhys.62.867} for standard books and reviews), generalizing the original harmonic oscillator coherent states (CSs) of the radiation field \cite{PhysRev.131.2766} and atomic CSs in quantum optics \cite{GilmorePhysRevA.6.2211}. CSs turn out to be good variational states, which faithfully reproduce the ground state energy of critical quantum systems undergoing a QPT, occurring in the thermodynamic limit and for some critical value $\lambda_c$ of a Hamiltonian control parameter (usually the two-body coupling constant). Here we shall extend the use of U$(N)$  CSs  from the most symmetric sector, where the ground state belongs, to other less symmetric permutation symmetry sectors $h$ or IRs, arising in the  $P$-fold tensor product decomposition. In each of these U$(N)$ IRs, the U$(N)$ CSs also describe the lowest-energy state, specially in the limit $P\to\infty$, where a generalized QPT occurs for a critical value $\lambda_c(h)$. This concept is called ``mixed permutation symmetry QPT'' and was first minted in \cite{nuestroPRE}.  Using the Lipkin-Meshkov-Glick (LMG) model as a paradigmatic example, we shall show that the Lieb-Mattis predictions still hold in the thermodynamic limit for $N=3$ fermion spinor components. We also discuss the $N=2$ component case for pedagogical reasons.

The organization of the article is as follows. In Section \ref{QHFsec} we comment on general models for interacting multicomponent fermions on a lattice, like $\rmu(N)$-invariant QHFs. We also review essential (Young tableau) diagrammatic notation to classify the different  $\rmu(N)$ IRs and permutation symmetry sectors that arise  in the $P$-fold tensor product Clebsch-Gordan decomposition, as well as the characterization of highest weight vectors.  In Section \ref{LiebMattisSec} we briefly review the Lieb-Mattis ordering theorem of electronic energy levels in terms of Young diagrams and their ``dominance" relations or pouring principle. Section \ref{cohsec}  is devoted to the important concept of CSs on  Grassmannian phase spaces and their operator expectation values. In Section \ref{LMGsec} we exemplify our construction with a generalization of the traditional  LMG Hamiltonian model for $P$ interacting  identical two-level particles/atoms   to $N$ levels, focusing on the cases $N=2$ and $N=3$. We describe QPTs occurring for some critical values of the control parameters inside each permutation symmetry sector for $P\to\infty$ and prove the Lieb-Mattis theorem in this limit.
Section \ref{conclusec} is left for conclusions and \ref{app} for an appendix.

\section{A model of multicomponent fermionic mixtures on a lattice}\label{QHFsec}

We consider a system of  $P$ fermions with $N$ spinor components/species. We distribute those fermions on a (planar) lattice with $L$ lattice sites. For electrons on a planar sample of area $A$ in a perpendicular magnetic field $B$, we could also think of $L=A/A_0$ as the number of Landau sites of area $A_0=2\pi\ell_B$ with $\ell_B=\sqrt{\hbar/(eB)}$ the magnetic length. We consider a typical Hamiltonian model describing pairing correlations in a second quantized form
\begin{multline}
H=\sum_{\alpha=1}^{L}\sum_{i=1}^{N}\epsilon_i c_{i}^\dag(\alpha)c_{i}(\alpha)\\
-
\sum_{\langle\alpha,\beta\rangle}\sum_{i,j,k,l=1}^N
\lambda_{ijkl} c_{i}^\dag(\alpha) c_{j}(\alpha) c_{k}^\dag(\beta) c_{l}(\beta)\label{ham1}
\end{multline}
where $c_{i}^\dag(\alpha)$ [$c_{i}(\alpha)$] creates [destroys] a fermion with spinor component $i=1,\dots,N$ at the Lattice site $\alpha=1,\dots,L$,  $\epsilon_i$ denote splitting energies and  $\lambda$ is the strength of the two-body  residual interactions  which scatter
pairs of particles across near-neighbor lattice sites $\langle\alpha,\beta\rangle$. Defining the $\rmu(N)$-spin operators 
\begin{equation}
S_{ij}(\alpha)=c^\dag_{i}(\alpha)c_{j}(\alpha),\:\: i,j=1,\dots,N,\:\: \alpha=1,\dots, L,\label{UNspinop}
\end{equation}
with commutation relations
\begin{equation}\left[{S}_{{ij}}(\alpha),{S}_{{kl}}(\beta)\right]=\delta_{\alpha\beta}(\delta _{{jk}} {S}_{{il}}(\beta) -\delta _{{il}} {S}_{{kj}}(\beta)),
\end{equation}
the Hamiltonian \eqref{ham1} can be rewritten as 
\begin{equation}
H=\sum_{\alpha=1}^{L}\sum_{i=1}^{L}\epsilon_i S_{ii}(\alpha)-\sum_{\langle\alpha,\beta\rangle}\sum_{i,j,k,l=1}^N
\lambda_{ijkl} S_{ij}(\alpha)S_{kl}(\beta)\label{hamUL}\:.
\end{equation}
For two-body exchange interactions ($\epsilon_i=0, \lambda_{ijkl}=\mathcal{J}\delta_{il}\delta_{jk}$), the Hamiltonian adopts the form of a $\rmu(N)$-invariant QHF, with $\mathcal{J}$ the exchange coupling constant. For $N=2$ (spin) fermion  components, $\mathcal{J}$  is the spin stiffness for the XY Heisenberg model for magnetic interactions between adjacent $\langle\alpha,\beta\rangle$ dipoles, which tend to be aligned (lower energy) when $\mathcal{J}>0$ (ferromagnetic case). Given the Fourier transform $ \mathcal{S}_{ij}(q)=\sum_{\alpha=1}^\flux e^{\ic q\alpha} S_{ij}(\alpha)$,  the low momentum $q\simeq 0$ ground state excitations are described by the collective operators 
$\mathcal{S}_{ij}(0)=\sum_{\alpha=1}^\flux S_{ij}(\alpha)$. Note that collective operators are  invariant under site permutations $\alpha\leftrightarrow\alpha'$. The kind of IRs of $\rmu(N)$ related to lattice site translation invariant configurations  are those described by rectangular Young/Ferres diagrams of $M$ rows and $\flux$ columns 
\begin{equation}
	 [\flux^M]= M
	\Bigg\{
	\overbrace{
		\begin{gathered}
			\begin{ytableau}
				~ &...&~\\
				:& : & : \\
				~&...&~
			\end{ytableau}
		\end{gathered}
	}^{\flux}
\end{equation}
where $M$ is the number of fermions per lattice site (also called \emph{filling factor} when talking about Landau sites), which is restricted by $M\leq N$ according to the Pauli exclusion principle for $N$-component fermions on a lattice site $\alpha$. We use the shorthand 
$[L,\stackrel{M}{\dots},L]=[L^M]$ to write Young diagrams with $M$ rows of equal length $L$. 
The corresponding $P=ML$ particle fermion mixture $[L^M]$ is symmetric under lattice site (row) permutations (that is, lattice sites become indistinguishable/bosonized), and antisymmetric (fermionic) under (column) permutation of the $M$ fermions on a given lattice site $\alpha$. If we order splitting energies by $\epsilon_i<\epsilon_j, i<j$, the ground state  for low residual interaction strength $\lambda\ll 1$ will be approximately 
 ($|0\ra$ denotes the Fock vacuum) 
\begin{equation} 
|[L^M],0\ra\equiv\Pi_{\alpha=1}^{\flux}\Pi_{i=1}^M c_i^\dag(\alpha)|0\ra,\label{GS}
\end{equation}
which fills all $\flux$ lattice/Landau sites with the first $M$ fermion species (components/flavors) $i=1,\dots,M\leq N$. The vector $|[L^M],0\ra$ is also called the ``highest-weight'' (HW) vector; see later on this section for a general definition of HW vectors $|h,0\ra$ for general IRs $h$. This ground state spontaneously breaks the $\rmu(N)$ symmetry since a general unitary transformation mixes the first $M$  occupied internal levels with the $N-M$ unoccupied ones. The ground state $|[L^M],0\ra$ is  still invariant under the stability subgroup $\rmu(M)\times \rmu(N-M)$ of transformations among the $M$ occupied  and 
the $N-M$ unoccupied internal levels, respectively.  Therefore, the $\rmu(N)$  transformations that do not leave  $|[L^M],0\ra$ invariant create \emph{coherent excitations} above it. For higher two-body interaction strength $\lambda$, the ground state is well approximated by a coherent excitation  
above the HW state $|[L^M],0\ra$, especially at the thermodynamic limit $L\to\infty$. This is the Gilmore procedure \cite{Gilmorebook}. We shall study CSs in more detail in section \ref{cohsec}.

We want to explore beyond fully site-translation invariant configurations described by Young diagrams $[L^M]$. In fact, the Hilbert space of a $\rmu(N)$ QHF with $\flux$ lattice sites at integer filling factor $M$  is the $\tbinom{N}{M}^\flux$-dimensional $\flux$-fold 
tensor product vector space 
\begin{equation}
	\mathcal H_{N}^{\otimes\flux}[1^M]=\bigotimes_{\alpha=1}^\flux\mathcal H_{N}^\alpha[1^M],
\end{equation}
where $\mathcal H_{N}^\alpha[1^M]$ denotes  the $D[1^M]=\tbinom{N}{M}$-dimensional carrier Hilbert space at site $\alpha$ of the fully antisymmetric IR  of $\rmu(N)$ described by the Young diagrams of shape $[1^M]$, 
that is, with $M$ boxes on a single column. Basis vectors  of $\mathcal H_{N}^\alpha[1^M]$ are the  $M$-particle  Slater determinants represented in Fock and  Young tableau notation as
\begin{equation}
\Yvcentermath1  \Pi_{\mu=1}^M c_{i_\mu}^\dag(\alpha)|0\ra=\young({{i_1}},:,{{i_M}}) 
\end{equation}
and obtained by filling out columns of the corresponding Young diagram with components $i_\mu\in\{1,\dots,N\}$ in strictly increasing order $i_1<\dots <i_M$. One can see that indeed there are exactly $D[1^M]=\tbinom{N}{M}$ different arrangements of this kind, which coincides with the dimension of $\mathcal H_{N}^\alpha[1^M]$. Likewise, the $P(=M\flux)$-particle ground state  in \eqref{GS} is associated to the Young tableau
\begin{equation}
\Yvcentermath1 |[L^M],0\ra=\young(1\mydots 1,:::,M\mydots M)\label{HWLM}
\end{equation}
belonging to the carrier Hilbert space $\mathcal H_N[\flux^M]$ of the rectangular IR $[\flux^M]$ with dimension 
\begin{multline}
D{[\flux^M]}=\frac{\prod_{i=N-M+1}^{N}\binom{i+\flux-1}{i-1}}{\prod_{i=2}^M\binom{i+\flux-1}{i-1}}\\
\stackrel{M=1}{\longrightarrow}\binom{\flux+N-1}{\flux}\stackrel{N=2}{\longrightarrow} L+1=2s_0+1,\label{dimLM}
\end{multline}
where we have highlighted the case of $L$ symmetric quNits ($M=1$) and qubits ($N=2$ components and total spin $s_0=L/2$).

The $\flux$-fold tensor product space $\mathcal H_{N}^{\otimes\flux}[1^M]$ can be graphically represented in  Young diagram notation as
\begin{equation}
\Yvcentermath1  \\[3pt] M\Bigg\{\young(\quad,:,\quad)  \:\:\otimes\:\: \stackrel{\flux\, \mathrm{times}}{\dots}  \:\:\otimes\:\:\young(\quad,:,\quad) \quad \leftrightarrow \quad  
[1^M]^{\otimes\flux}=[1^M]\otimes\stackrel{\flux}{\dots}\otimes [1^M]\,.\label{tensorprod}
\end{equation}
This tensor product representation of $\rmu(N)$ is reducible and it decomposes into a direct sum of IRs
of different Young diagram shapes $h=[h_1,\dots,h_N]$  of 
$P=h_1+\dots h_N$ boxes/particles ($h$ is also called a partition of $P$)
\begin{equation}\label{youngdiagram}
	\overbrace{
		\begin{gathered}
			\begin{ytableau}
				~ &...&...&...&...&...&~\\
				: &:& : & : &:\\
				~&...&~
			\end{ytableau}
		\end{gathered}
	}^{h_1}
\end{equation}
with  $h_1\geq \dots \geq h_N$ and $h_i$ the number of boxes in row $i=1,\dots,N$. As already commented, one sometimes uses the shorthand 
$h=[L,\stackrel{M}{\dots},L,0,\dots ,0]=[L^M]$, obviating zero-box rows.

The best known example is perhaps the coupling of $P=L$ spin-1/2 particles ($N=2$, $M=1$), for which the dimensions of the Clebsch-Gordan 
decomposition series of the $L$-fold tensor product $[1]^{\otimes L}$ is given by the Catalan's triangle formula
\begin{equation}
[1]^{\otimes\flux}=\bigoplus_{k=0}^{\left \lfloor{L/2}\right \rfloor}\mathcal{M}_k[L-k,k],  \label{Catalan}
\end{equation}
with its respective dimension formula
\begin{align}
	2^{L}=&\,\sum_{k=0}^{\left \lfloor{L/2}\right \rfloor}(L+1-2k)\mathcal{M}_k,\nn\\ \mathcal{M}_k=&\,\frac{L+1-2k}{L+1}\binom{L+1}{k},
\end{align}
where $\left \lfloor{L/2}\right \rfloor$ is the integer floor function. That is, the spin $s_k=\frac{L}{2}-k$ irreducible representation $[L-k,k]$ appears 
with multiplicity $\mathcal{M}_k$ and dimension $D[L-k,k]=L+1-2k=2s_k+1$. Note that the fully symmetric representation $k=0$ ($s_0=L/2$) always appears with multiplicity $\mathcal{M}_0=1$. For example, for $P=L=2$ spin-$1/2$ particles, we have $[1]^{\otimes 2}=[1^2]\oplus [1,1]$, that is: spin $s_0=1$ triplet plus spin $s_1=0$ singlet.

For $N>2$ we do not have such closed formulas and the technical complexity increases. For example,  the Clebsch-Gordan decomposition of a tensor product of $\flux=2$ IRs of $\rmu(4)$ of shape $[1^2]$ (filling factor $M=2$)  is represented by the following Young diagrams
\begin{multline}
\Yvcentermath1   \young(\quad,\quad) \:\:\otimes\:\: \young(\quad,\quad)\:\:=\:\:\young(\quad\quad,\quad\quad)\:\:\oplus\:\: \young(\quad\quad,\quad,\quad)\:\:\oplus\:\:\young(\quad,\quad,\quad,\quad)  \quad \\
\leftrightarrow \quad  [1^2]\otimes[1^2]=[2^2]\oplus [2,1^2]\oplus [1^4].\label{lambda2nu2}
\end{multline}

To obtain the IR $h$ basis vectors (Young tableau configurations), Young diagrams columns must be  filled with fermion components $i_\mu\in\{1,\dots,N\}$ in strictly increasing order $i_1<i_2<\dots$ and rows in a non-decreasing order $i_1\leq i_2\leq \dots $ Counting up the number of resulting different Young tableaux, one can see that the corresponding dimensions for the tensor product decomposition \eqref{lambda2nu2} for $N=4$ is $6\times 6=20+15+1$. The dimension formula for a general IR $h=[h_1,\dots,h_N]$ is given by
\begin{equation}
 D[h]=\frac{\prod_{i<j}(h_i-h_j+j-i)}{\prod_{i=1}^{N-1}i!},\label{dimh}
\end{equation}
which comes from the Hook length formula \cite{Barut}, of which the dimension \eqref{dimLM} is a particular case.

The weight $w=[w_1,\dots,w_N]$ of a given Young tableaux (basis vector) counts the number of fermions $w_i$ with component/flavor $i$; that is, $w_i$ is the population/occupation  of the internal level $i$. A Young tableaux with weight $w$ is said to be of lower weight than other of weight $w'$ if the first non-vanishing component of $w-w'$ is positive. Therefore, the HW vector $|h,0\ra$ is obtained as the Young tableaux of shape $h=[h_1,h_2,\dots,h_N]$  
\begin{equation}
|h,0\ra=\Yvcentermath1 	
		\begin{gathered}
			\begin{ytableau}
				1&...&...&...&...&...&...&1\\
				2&...&...&...&...&2\\
				: &:& : & : \\
				N&...&N
			\end{ytableau}
		\end{gathered}
\end{equation}
which arises when filling out row $j$ with $h_j$ fermion components $j$. In more physical terms, HW vectors are obtained by populating lower fermion components $i=1,\dots,N$ when filling out the corresponding Young diagram according to the general rules mentioned above. One example is already given in \eqref{HWLM}. For $N=2$ spin components $i=1,2$ and $P$ fermions, the HW vector of $h=[h_1,h_2]=[(2s+P)/2,(P-2s)/2]$ is 
\begin{equation}
 |[h_1,h_2],0\ra=\Yvcentermath1  \young(1\mydots\mydots\mydots\mydots\mydots  1,2\mydots\mydots\mydots 2)
\end{equation}
which has total spin $s=(h_1-h_2)/2$ and highest spin third component $m=s$; in general, $m$ coincides with the half difference between the populations/occupations of flavors 1 and 2, i.e. $m=(w_1-w_2)/2$ in terms of weights. Angular momentum basis vectors are usually written as $\{|s,m\ra, m=-s,\dots,s\}$ so that, in ``Dicke'' notation, the highest-weight vector is $|[h_1,h_2],0\ra=|s,s\ra$.  Rising (resp. lowering) $\rmu(N)$-spin ladder operators $S_{ji}, j<i$ (resp. $S_{ji}, j>i$) are meant to rise (resp. lower) weight. For $N=2$ one recovers the usual angular momentum operators as $J_3=(S_{11}-S_{22})/2, J_+=S_{12}$ and $J_-=S_{21}$.

As already commented before \eqref{GS}, for low residual interaction strength $\lambda\ll 1$, and splitting energies ordered as $\epsilon_i<\epsilon_j, i<j$, HW vectors $|h,0\ra$ are designed to be the lowest energy states inside a  given permutation symmetry sector $h$. Let us see that there is also a ``dominance order'' $h\succeq h'$ between different permutation symmetry sectors $h$ and $h'$.

\section{Lieb-Mattis ordering theorem of electronic energy levels}\label{LiebMattisSec}

The set of Young diagrams arising in the Clebsch-Gordan decomposition of the $L$-fold tensor product \eqref{tensorprod} can be ``energetically ordered'' in the following sense. 
The procedure is based on the so called ``dominance order'' $\succeq$, or ``pouring principle'', implicit in the  Lieb-Mattis theorem 
\cite{Lieb-Mattis_PR1962} (see also \cite{Decamp_PRR2020} for some recent applications/generalizations). More precisely, given two Young diagrams $h$ and $h'$ of $P$ particles/boxes, it is said that $h$ ``dominates''  $h'$ ($h\succeq h'$) [or that $h'$ ``precedes'' $h$ ($h'\preceq h$), or that $h'$ can be poured into $h$ ($h'\stackrel{p}{\to}h$)] if 
\begin{multline}\label{DominanceIneq}
[h_1,\dots,h_N]\succeq [h_1',\dots,h_N']\\[1em]
\Leftrightarrow h_1+\dots+h_k\geq h_1'+\dots+h_k' \quad \forall k\in\{1,\dots,N\}\,.
\end{multline}
Intuitively, it means that one can go from $h$ to $h'$ by moving a certain number of boxes from
upper rows to lower rows, so that $h$ is ``more symmetric'' than $h'$. For example, one can see that the rectangular Young diagram
 $[2^2]$ in \eqref{lambda2nu2}] dominates over the rest of Young diagrams ( $[2^2]\succeq [2,1^2]\succeq [1^4]$) arising in the Clebsch-Gordan decomposition of $[1^2]\otimes[1^2]$. In general, one can prove 
 the following proposition (see \cite{sym14050872} for a proof).

\begin{prop}\label{rectdomin} All Young diagrams arising in the Clebsch-Gordan direct sum decomposition of the $\flux$-fold 
tensor product \eqref{tensorprod} can be poured into the rectangular Young diagram of shape $[\flux^M]$. That is, $[\flux^M]$ dominates over the rest of Young diagrams 
 arising in the Clebsch-Gordan decomposition of $[1^M]^{\otimes L}$
\end{prop}

Note that, in general, not all arbitrary $P$-particle Young diagrams $h$ and $h'$ can be compared, as either $h\succeq h'$ or $h\preceq h'$ not always hold for $P>5$ \cite{Decamp_PRR2020}. For instance, $h=[4,1,1]$ and $h'=[3,3,0]$ fulfill $h_1>h_1'$ but $h_1+h_2<h_1'+h_2'$, hence neither $h\succeq h'$ nor $h\preceq h'$ is true. This is why the set of $P$-particle Young diagrams is only partially ordered.

Lieb-Mattis theorem \cite{Lieb-Mattis_PR1962}  states that, under general conditions on the symmetric Hamiltonian of the system (i.e., non ``pathologic'' potentials), if $h'$ can be poured into  $h$ (that is, $h\succeq h'$) then $E(h)<E(h')$  with $E(h)$ the lowest-energy eigenvalue inside each IR $h$ of $\rmu(N)$. Actually, the original Lieb-Mattis theorem is restricted to $N=2$ spin components, so that IRs $h=[h_1,h_2]$ of $\rmu(2)$ with $P=h_1+h_2$ particles can be labelled by $h=[(2s+P)/2,(P-2s)/2]$, with $s=(h_1-h_2)/2$ the total spin; the most symmetric Young diagram corresponds to $h_1=P=2s, h_2=0$, to where the ground state of the $\rmu(2)$ QHF at filling factor $M=1$ belongs. Remember that the degeneracy $\mathcal{M}_k$ of each $\rmu(2)$ IR of spin $s_k=\frac{L}{2}-k, k=0,\dots,\left \lfloor{L/2}\right \rfloor$, arising in the the Clebsch-Gordan 
decomposition series of  $[1]^{\otimes L}$, is given in \eqref{Catalan}, with $\mathcal{M}_0=1$. 
In general, for a  $\rmu(N)$ QHF at filling factor $M$, the ground state belongs to the carrier Hilbert space $\mathcal H_N[\flux^M]$ of the rectangular IR $[\flux^M]$ of  $\rmu(N)$  inside the total Hilbert space $\mathcal H_{N}^{\otimes\flux}[1^M]$.

We want to investigate whether or not the Lieb-Mattis theorem still holds in the thermodynamic limit $P\to\infty$ and how to deal with it. For this purpose, we shall introduce CSs associated to a given 
IR $h$ of $\rmu(N)$. When the residual interaction strength $\lambda$ is not low, coherent excitations $|h,Z\ra$ (see next section) above the HW state $|h,0\ra$ turn out to faithfully reproduce the energy of the lowest-energy eigenvector of a multiparticle Hamiltonian inside each permutation symmetry sector $h$. In the thermodymamical limit $P\to\infty$, CS Hamiltonian expectation values $\la h,Z|H|h,Z\ra$ capture the critical points for systems underlying a QPT; see e.g \cite{octavio,Romera_2014} for the case of the $N=2$ level LMG model and \cite{PhysRevA.92.053843,PhysRevA.94.013802} for 
a system of $P$ indistinguishable atoms of $N$ levels interacting dipolarly with
$\ell$ modes of an electromagnetic field.

Let us show how to construct CS and matrix elements of $\rmu(N)$-spin operators $S_{ij}$. Eventually we shall resort to the less technical cases, $N=2$ and $N=3$ fermion components, to exemplify the general case.

\section{U(N) Coherent  states and their operator expectation values}\label{cohsec}

Coherent states $|h,Y\ra$ are  defined as $\rmu(N)$ rotations/excitations $\mathcal{U}(Y)$ above a highest-weight vector $|h,0\ra$ as
\begin{equation}
|h,Y\ra=\mathcal{U}(Y)|h,0\ra=e^{\ic\sum_{i,j=1}^N y_{ij}S_{ij}}|h,0\ra,
\end{equation}
where $y_{ij}$ are complex coordinates fulfilling $y_{ji}=\bar{y}_{ij}$  since $S_{ij}^\dag=S_{ji}$ and $\mathcal{U}(Y)$ is a unitary transformation. 
The HW vector $|h,0\ra$ is always invariant (up to a phase) under the stability subgroup $\rmu(1)^N\subset \rmu(N)$ made of diagonal transformations $\exp(\sum_{i=1}^N y_{ii}S_{ii})$. Therefore, the $\rmu(N)$ transformations $\mathcal{U}(Y)$ that do effectively create coherent excitations above $|h,0\ra$ belong to the coset $\mathbb{F}_N=\rmu(N)/\rmu(1)^N$ (\emph{flag manifold}), which is parametrized by $Y=(y_{ij}), i>j$, arranged as a complex lower-triangular $N\times N$ matrix; therefore, the complex dimension of the phase space manifold $\mathbb{F}_N$ is  $\dim_\mathbb{C}(\mathbb{F}_N)=N(N-1)/2$. The stability subgroup of $|h,0\ra$ can be larger than $\rmu(1)^N$; for example, as we already commented after \eqref{GS}, the stability subgroup of $|[L^M],0\ra$ is $\rmu(M)\times \rmu(N-M)$, so that coherent excitations $|[L^M],Y\ra$ above $|[L^M],0\ra$ are labelled by $(N-M)\times M$ complex coordinates $y_{ij}, M+1\leq i\leq N, 1\leq j\leq M$ parametrizing the coset $\mathbb{G}^N_M=\rmu(N)/\rmu(M)\times \rmu(N-M)$, which we identify with a \emph{Grassmannian manifold}. For filling factor $M=1$, the Grassmannian manifold becomes $\mathbb{G}^N_1=\mathbb{C}P^{N-1}$ the complex projective space, which is the phase space of $P=L$ ``symmetric quNits'' \cite{QIP-2021-Entanglement}. In fact, for $N=2$ spinor components and filling factor $M=1$, the matrix $Y$ becomes a scalar complex coordinate $y=-\ic\frac{\theta}{2}e^{\ic\phi}$ on the sphere $\mathbb{S}^2=\mathbb{G}^2_1=\rmu(2)/\rmu(1)^2$ (the complex projective $\mathbb{C}P^1$ space in this case) 
given in terms of the spherical polar $\theta$ and azimuthal $\phi$ coordinates.

The fact that the HW vector $|h,0\ra$ is annihilated by any rising $\rmu(N)$-spin operator $S_{ij}, i<j$, together with the Baker-Campbell-Hausdorff-Zassenhaus factorization formula applied to the group $\rmu(N)$ (see e.g. \cite{Gazeaubook}),  allows one to write
\begin{align}
|h,Y\ra=&\,K_h(Z)e^{\sum _{j<i} z_{ij}S_{ij}}|h,0\ra,\nn\\
 Z=&\,\ic Y(Y^\dag Y)^{-1/2}\tan(Y^\dag Y)^{1/2},
\end{align}
where $K_h(Z)$ is a normalization factor to be discussed later. Abusing notation we will denote by $|h,Y\ra=|h,Z\ra$, in the hope that no confusion arises.

For $N=2$ spin components we recover spin $s=(h_1-h_2)/2$, $\rmu(2)$ or ``atomic''  CSs \cite{Gilmore} 
\begin{align}
|[h_1,h_2],z\ra=&\,(1+|z|^2)^{-s}e^{zS_{21}}|[h_1,h_2],0\ra,\nn\\ z_{21}=z=&\,\frac{y}{|y|}\tan|y|,\; y=y_{21}=\frac{\theta}{2}e^{\ic \phi},
\end{align}
where $S_{21}$ is the spin lowering operator $S_-$ and $z=\tan(\theta/2)e^{\mathrm{i}\phi}$ can be identified with the stereographic projection of the Bloch sphere $\mathbb{S}^2\ni (\theta,\phi)$ onto the complex plane $\mathbb{C}$. Spanning $e^{zS_{21}}|[h_1,h_2],0\ra=e^{zS_-}|s,s\ra$ in terms of Dicke states $\{|s,m\ra, m=-s,\dots,s\}$ we obtain a more familiar form of spin-$s$ CSs 
\begin{equation}
|[h_1,h_2],z\ra=(1+|z|^2)^{-s}\sum_{m=-s}^s\sqrt{\binom{2s}{s-m}}z^{s-m}|s,m\ra.
\end{equation}
The normalization factor $K_{[h_1,h_2]}(z)=(1+|z|^2)^{-s}$ for $\rmu(2)$ is generalized to $\rmu(N)$ as follows. We identify the ``length'' $(1+|z|^2)$ as the leading principal minor 
\begin{equation}
|Z^\dag Z|_1=1+\bar{z}z,\; Z=\left(
\begin{array}{cc}
 1 & 0  \\
 z  & 1 
\end{array}
\right).\label{GSU2}
\end{equation}
The same procedure applies for $N=3$ fermion components, where now we have two lengths  coming from the two upper-principal minors or order one and two
\begin{align}
	|Z^\dag Z|_1=&\,1+ \bar{z}_1z_1+\bar{z}_2z_2,\nn\\
	|Z^\dag Z|_2=&\,1+ \bar{z}_3z_3+ (\bar{z}_2-\bar{z}_1 \bar{z}_3)(z_2 -z_1  z_3 ),\nn\\
	Z=&\,\left(
	\begin{array}{ccc}
		1 & 0 & 0 \\
		z_1 & 1 & 0 \\
		z_2 & z_3  & 1 \\
	\end{array}
	\right). \label{lengths}
\end{align}
In general, for $\rmu(N)$, we will have $N-1$ lengths  coming from the $N-1$ upper-principal minors of $Z^\dag Z$, with $Z$ a lower-triangular complex $N\times N$ matrix. The normalization factor $K_h(Z)$ then turns out to be written as
\begin{multline}
K_h(Z)= |Z^\dag Z|_1^{(h_2-h_1)/2}|Z^\dag Z|_2^{(h_3-h_2)/2}\dots\\
	\dots |Z^\dag Z|_{N-1}^{(h_{N}-h_{N-1})/2}.\label{normcoh}
\end{multline}
An interesting function is the reproducing Bergman kernel (CS overlap)
\begin{align}
B_h(Z',Z)=&\,\frac{\langle h,Z'|h,Z\ra}{{K_h(Z')}K_h(Z)}\nn\\
=&\,|Z'^\dag Z|_1^{h_1-h_2}|Z'^\dag Z|_2^{h_2-h_3}\dots |Z'^\dag Z|_{N-1}^{h_{N-1}-h_N},
\end{align}
defined in terms of the $N-1$ upper-principal minors of $Z'^\dag Z$. It is holomorphic in $Z$ and antiholomorphic in $Z'$. For example, for $N=3$ we explicitly have 
\begin{multline}
B_{[h_1,h_2,h_3]}(Z',Z)=(1+ \bar{z}'_1z_1+\bar{z}'_2z_2)^{h_1-h_2}\\
\times(1+ \bar{z}'_3z_3+ (\bar{z}'_2-\bar{z}'_1 \bar{z}'_3)(z_2 -z_1  z_3 ))^{h_2-h_3}.
\end{multline}
Coherent state expectation values $\mathbb{s}_{ij}$ of the basic symmetry operators $S_{ij}$ can be easily computed through derivatives of the Bergman kernel as
\begin{equation}
\mathbb{s}_{ij}=\langle h, Z|S_{ij}|h, Z\rangle=\frac{\mathbb{S}_{ij}B_h(Z,Z)}{B_h(Z,Z) }.\label{sijsymb}
\end{equation}
where $\mathbb{S}_{ij}$ denotes the  differential representation  of the $\rmu(N)$-spin $S_{ij}$ operators \eqref{UNspinop} on anti-holomorphic functions $\psi(Z^\dag)=\langle h,Z|\psi\ra/K_h(Z)$. This differential representation can be obtained in several ways; one is by using operator commutation properties like $[A,e^{zB}]=\pm z\partial_ze^{zB}$ for $[A,B]=\pm B$, etc. The procedure is in general quite technical and we shall present here the results for $N=2$ and $N=3$ fermion components. For example, for $\rmu(2)$ we have 
\begin{align}
(\mathbb{S}_{ij})=&\,\left(\begin{array}{ccc}h_1- \bar{z}\partial_{\bar{z}} & \hspace{1mm} & \partial_{\bar{z}}\\ 
(h_1-h_2)\bar{z}-\bar{z}^2\partial_{\bar{z}} & & h_2+\bar{z}\partial_{\bar{z}}\end{array}\right),\nonumber\\
(\mathbb{s}_{ij})=&\,\frac{1}{1+|z|^2}\left(\begin{array}{ccc}h_1+h_2|z|^2& \hspace{1mm} & (h_1-h_2)z\\ (h_1-h_2)\bar{z}& & h_2+h_1|z|^2\end{array}\right). \label{sijN2}
\end{align}
For $\rmu(3)$, the differential representation of $\rmu(3)$-spin operators is
\begin{align}
	\mathbb{S}_{11}=&\,h_1- \bar{z}_2  \partial_{\bar{z}_2}-\bar{z}_1   
	\partial_{\bar{z}_1}\,,\nn \\
	\mathbb{S}_{21}=&\,\bar{z}_1   (h_1-h_2)-(\bar{z}_2 -\bar{z}_1  \bar{z}_3 ) \partial_{\bar{z}_3}-\bar{z}_1   \left(\bar{z}_2  
	\partial_{\bar{z}_2}+\bar{z}_1 
	\partial_{\bar{z}_1}\right) ,\nn\\
	\mathbb{S}_{31} =&\, (h_1-h_3)\bar{z}_2 + (h_3-h_2)\bar{z}_1  \bar{z}_3 -
	\bar{z}_3  (\bar{z}_2 -\bar{z}_1  \bar{z}_3 ) \partial_{\bar{z}_3} ,\nn\\
	\mathbb{S}_{12} =&\, \partial_{\bar{z}_1},\nn\\
	\mathbb{S}_{22} =&\, h_2 +\bar{z}_1   \partial_{\bar{z}_1}-\bar{z}_3  \partial_{\bar{z}_3} ,\nn\\ 
	\mathbb{S}_{32} = &\,(h_2-h_3) \bar{z}_3 -
	\bar{z}_3^2 \partial_{\bar{z}_3}+\bar{z}_2  \partial_{\bar{z}_1},\nn\\
	\mathbb{S}_{13} =&\, \partial_{\bar{z}_2},\nn\\
	 \quad\mathbb{S}_{23} \,=&\, \partial_{\bar{z}_3}+\bar{z}_1   \partial_{\bar{z}_2} ,\nn\\
	 \quad\mathbb{S}_{33} =&\, h_3+\bar{z}_3  \partial_{\bar{z}_3}+\bar{z}_2  \partial_{\bar{z}_2}\,,
	\label{difrelU3}
\end{align}
and their CS expectation values 
\begin{align}
\mathbb{s}_{11}=&\,  \frac{h_1}{|Z^\dag Z|_1} +  \frac{h_2|z_1 +z_2  \bar{z}_3 |^2}{|Z^\dag Z|_1|Z^\dag Z|_2}+  \frac{ h_3|z_2 -z_1  z_3|^2}{|Z^\dag Z|_2}, \nonumber \\
\mathbb{s}_{22}=& \, \frac{h_1|z_1|^2}{|Z^\dag Z|_1}   + \frac{ h_2|1-z_1  \bar{z}_2  z_3 +z_2  \bar{z}_2|^2}{ |Z^\dag Z|_1|Z^\dag Z|_2}+  \frac{h_3 |z_3|^2}{|Z^\dag Z|_2},\nonumber\\
\mathbb{s}_{33}=& \,  \frac{ h_1 |z_2|^2  +  h_2(1+|z_1|^2)}{|Z^\dag Z|_1}+\frac{ h_3- h_2}{|Z^\dag Z|_2},\nonumber\\
\mathbb{s}_{12}=& \,    \frac{(h_1-h_2) z_1}{|Z^\dag Z|_1}+ \frac{( h_2-h_3)\bar{z}_3 (z_2 -z_1  z_3 )}{|Z^\dag Z|_2}, \nonumber \\
\mathbb{s}_{13}=& \,  \frac{( h_1- h_2)z_2  }{|Z^\dag Z|_1}+\frac{( h_2- h_3) (z_2 -z_1  z_3 )}{|Z^\dag Z|_2}, \nonumber \\
\mathbb{s}_{23}=& \,  \frac{( h_1- h_2)\bar{z}_1  z_2  }{|Z^\dag Z|_1}+\frac{( h_2- h_3)z_3  }{|Z^\dag Z|_2}
 \label{CSEV}
 \end{align}
and $\mathbb{s}_{ij}=\bar{\mathbb{s}}_{ji}$ for the reminder. Coherent state expectation values of higher powers of $\rmu(N)$-spin operators $S_{ij}$ can also be easily computed by repeated differentiation of the Bergman kernel. For example, 
for quadratic powers we have
\begin{equation}
\langle h, Z|S_{ij}S_{kl}|h, Z\rangle=\frac{\mathbb{S}_{ij}\left(\mathbb{S}_{kl} B_h(Z,Z)\right)}{B_h(Z,Z) }.
\end{equation}
This calculation will be useful when evaluating two-body interaction terms in the Hamiltonian \eqref{hamUL}. In the thermodynamic/semi-classical limit $P\to\infty$ this calculation simplifies since   quantum fluctuations disappear  and we have 
\begin{equation}
\lim_{P\to\infty}\frac{ \langle h, Z|S_{ij}S_{kl}|h, Z\rangle}{\langle h, Z|S_{ij}|h, Z\rangle\langle h, Z|S_{kl}|h, Z\rangle}= 1,\label{nofluct}
\end{equation}
which means that, in this limit, CS expectation values of quadratic powers $S_{ij}S_{kl}$ can  be simply written as the product $\mathbb{s}_{ij}\mathbb{s}_{kl}$.

As we have already said, for higher two-body interaction strength $\lambda\not=0$, the lower-energy state inside each permutation symmetry sector $h$ turns out to be  well approximated by a coherent excitation $|h,Z\ra$ above the HW state $|h,0\ra$, especially at the thermodynamic limit $P\to\infty$. These CSs play a fundamental role in the study of the low-energy long-wavelength semi-classical ($\flux\to\infty$) dynamics of $\rmu(N)$ quantum Hall ferromagnets  described by nonlinear sigma models  which target space is the Grassmannian  $\mathbb{G}^N_M$ phase space manifold \cite{AffleckNPB257,AffleckNPB265,AffleckNPB305,Sachdev,Sachdev2,Arovas,CALIXTO1}. In the next section we shall use these CS expectation values  to compute the energy surface $\la h,Z|H|h,Z\ra$ of our Hamiltonian $H$ for each permutation symmetry sector $h$ of the system.

\section{LMG model and phase diagram in each permutation symmetry sector}\label{LMGsec}

To test and describe the Lieb-Mattis theorem in the thermodynamic limit, we shall make use of a generalization of the traditional $N=2$-level LMG model to arbitrary $N$ levels. The ubiquitous LMG Hamiltonian model acquires multiple forms. The closest  to our construction  \eqref{hamUL}, with filling factor $M=1$, is the one of  an anisotropic XY Ising model 
\begin{multline}
H_{XY}=  \epsilon \sum_{\alpha=1}^L \sigma_z(\alpha)
+  \sum_{\alpha<\beta} \lambda_x \sigma_x(\alpha)  \sigma_x(\beta) \\
 +   \sum_{\alpha<\beta} \lambda_y \sigma_y(\alpha)  \sigma_y(\beta)\,,
\label{Hlmgeneralpauli}
\end{multline}
in an external perpendicular magnetic field $\epsilon$ with infinite-range constant
interactions $\lambda_{x,y}$, and $\vec{\sigma}(\alpha)=/\sigma_x(\alpha),\sigma_y(\alpha),\sigma_z(\alpha))$ the Pauli matrices at lattice site $\alpha$. In terms of collective $U(2)$ angular momentum operators 
$\vec{J}= \sum_{\alpha=1}^L \vec{\sigma}(\alpha)$, the two-level LMG Hamiltonian acquires the form \cite{lipkin1,lipkin2}:
\begin{equation}
H_{XY} = \epsilon J_z+\frac{\lambda_1}{2}(J_+^2+J_-^2)+\frac{\lambda_2}{2}(J_+J_-+J_-J_+)\:,
\label{hamU2}
\end{equation}
with $\lambda_{1,2}$ a function of $\lambda_{x,y}$. The $\lambda_1$  term causes the annihilation of particles at one of the two levels and the creation of particles at the other  level, and the $\lambda_2$ term results in the scattering of one particle upwards while simultaneously scattering another particle downwards. The total number of particles $P=L=2j$  and $\vec{J}^2=j(j+1)$  are conserved. Permutation and U(2)  
symmetries reduce the size of the largest matrix to be diagonalized from $2^P$ to $P+1=2j+1$.

We extend this model to $N$-level atoms with Hamiltonian density
\begin{equation}
H_N=\frac{\epsilon}{P}(S_{NN}-S_{11})-\frac{\lambda}{P(P-1)}\sum_{i\not=j=1}^{N}S_{ij}^2,\label{Hd}
\end{equation}
described by collective $\rmu(N)$ operators $S_{ij}=\sum_{\alpha=1}^L S_{ij}(\alpha)$. For simplicity, we have discarded interactions of fermions with the same spinor component [i.e. we have disregarded terms of the type $S_{ij}S_{ji}+h.c.$ like the ones accompanying $\lambda_2$ in \eqref{hamU2}] and kept equal interactions for fermions with different spinor components/flavors [i.e. we have maintained $S_{ij}^2$  like the ones accompanying $\lambda_1$ in \eqref{hamU2}] so that the parity $e^{\ic \pi S_{ii}}$ of the population $S_{ii}$ of each flavor $i$ is conserved. We have also disposed splitting energies $\epsilon_i$ equally spaced $\epsilon_i=\epsilon$ and symmetrically about $i=(N+1)/2$. Since we are interested in the thermodynamic limit $P\to\infty$, we have made the  Hamiltonian density \eqref{Hd} an intensive quantity by dividing one-body interactions (those proportional to $S_{ij}$) by the total number of particles $P$, and two-body interactions (those proportional to $S_{ij}^2$) by the number of particle pairs $P(P-1)$. In the following we shall restrict ourselves to $N=2$ and $N=3$ spinor components ($N$ atom levels in the usual LMG jargon), for simplicity. We shall take $\epsilon=1$, which is equivalent to measure the energy and $\lambda$ in $\epsilon$ units.

\subsection{$N=2$ fermion spinor components}

The CS expectation value of the Hamiltonian density $H_2$ in each permutation symmetry sector $h=[h_1,h_2]=[P/2+s,P/2-s]$ of $P=h_1+h_2$ particles and total spin $s=(h_1-h_2)/2$ is
\begin{equation}
\la h,z|H_2|h,z\ra=\frac{1}{P}(\mathbb{s}_{22}-\mathbb{s}_{11})-\frac{\lambda}{P(P-1)}\sum_{i\not=j=1}^{2}\mathbb{s}_{ij}^2,
\end{equation}
where $s_{ij}$ is defined in \eqref{sijsymb} and its explicit expression is given in \eqref{sijN2}. The 
so called \emph{energy surface} is defined in the thermodynamic limit
\begin{align}
E_\mu^z(\lambda)=&\,\lim_{P\to\infty}\la h,z|H_2|h,z\ra\nn\\
=&\,(2\mu-1)\frac{|z|^4-1-\lambda(2\mu-1)(z^2+\bar{z}^2)}{(1+|z|^2)^2},
\end{align}
where we have defined the fraction $\mu=h_1/P=\frac{1}{2}+\frac{s}{P}$. Note that $\mu\in [\frac{1}{2},1]$ is a continuous (non discrete) parameter in the thermodynamic limit. Note also that $h$ dominates $h'$ ($h\succeq  h'$) iff $\mu>\mu'$ or $s>s'$. Let us see whether the Lieb-Mattis theorem still holds in the thermodynamic limit. Applying Gilmore's minimization procedure \cite{Gilmore}, the lowest-energy inside each permutation symmetry sector $\mu$ turns out to be
\begin{equation}
E^{(0)}_\mu(\lambda)=\min_{z\in\mathbb{C}}E_\mu^z(\lambda)=\left\{\begin{array}{lr} 1-2\mu,&\lambda< \frac{1}{2\mu-1},\\ -\frac{1+\lambda^2(1-2\mu)^2}{2\lambda}&\lambda\geq \frac{1}{2\mu-1},\end{array}\right. 
\end{equation}
with minimal coordinates
\begin{equation}
z_0^{\pm}=\left\{\begin{array}{lr} 0,&\lambda< \frac{1}{2\mu-1},\\ \pm\sqrt{\frac{\lambda(2\mu-1)-1}{\lambda(2\mu-1)+1}}&\lambda\geq \frac{1}{2\mu-1}.\end{array}\right.
\end{equation}
The fact that there are two ground states $|h,z_0^\pm\ra$ with the same energy $E^{(0)}_\mu(\lambda)$ is related to a spontaneous breakdown of the parity symmetry of the Hamiltonian $H_2$ in the thermodymamical limit.

We identify $\lambda_c(\mu)=1/(2\mu-1)$ as a critical value of the interaction strength control parameter for which a second order QPT inside the permutation symmetry sector $\mu$ takes place. This concept was firstly introduced by us in \cite{nuestroPRE} and termed ``mixed symmetry'' QPT (MSQPT). Standard QPTs are usually associated to the totally symmetric permutation symmetry sector $\mu=1, s=P/2$, for which $\lambda_c(1)=1$. MSQPTs extend this concept to general $\mu\in [\frac{1}{2},1]$ (general $h=[h_1,h_2]$). Other extensions of the QPT concept appear in the literature, but they still refer to the ground state (fully symmetric $\mu=1$) sector;  see e.g. \cite{ESQPT,relano} for the concept of ``Excited State Quantum Phase Transitions'' (ESQPTs) when excited states above the ground state are also analyzed.

Since
\begin{equation}
\partial_\mu E^{(0)}_\mu(\lambda)=\left\{\begin{array}{lr} -2,&\lambda< \lambda_c,\\ -2\lambda(2\mu-1)&\lambda\geq \lambda_c,\end{array}\right. 
\end{equation}
is non positive for $\mu\in [\frac{1}{2},1]$, the energy density $E^{(0)}_\mu$ is a decreasing function of $\mu$, and therefore of the spin $s$, in the thermodynamic limit. Thus, Lieb-Mattis theorem also applies in this limit. The case $\mu=1/2$ corresponds to the spin singlet $s=0$ and does not display a MSQPT since $\lambda_c(1/2)\to\infty$.


\subsection{$N=3$ fermion components}

In order to take the limit $P=h_1+h_2+h_3\to \infty$, we shall define the two proportions  $\mu, \nu$ as
\begin{align} h_3=\nu P, \quad h_2=&\,(1-\mu-\nu)P, \quad h_1=\mu P,  \nn\\
 \forall\, \nu\in[0,\tfrac{1}{3}],& \quad \mu\in[\tfrac{1}{2}(1-\nu),1-2\nu]. 
\label{mu_nu_def}
\end{align}
Note that the set  $(\mu,\nu)$ is dense in the corresponding intervals as $P\to\infty$. The parametrization in Eq. \eqref{mu_nu_def} has been chosen differently to others defined previously \cite{nuestroPRE}, since these expressions become more useful for proving the Lieb-Mattis theorem for the 3-level LMG model in the thermodynamic limit.

Following the same procedure as for the case $N=2$, we compute the energy surface in the thermodynamic limit using the LMG Hamiltonian \eqref{Hd} and the CS expectation values for U(3) in \eqref{CSEV}. In this case, the bulky expression of the energy surface $E^{Z}_{\mu,\nu}(\lambda)$ and the minimal coordinates $Z^{(0)}$ can be found in  Appendix \ref{app}. Here we proceed by providing just the final expression  of the minimal energy surface inside each mixed symmetry sector $(\mu,\nu)$ given by the expression

\begin{widetext}
\begin{equation}
	E_{\mu,\nu}^{(0)}(\lambda)=\left\{  
	\begin{array}{ll}
		\left. 
		\begin{array}{ll}
			-\mu +\nu,  &\quad 0\leq \lambda < \frac{1}{2-2\mu-4 \nu } \\
			-\frac{4 \lambda ^2 (\mu+2 \nu -1)^2+4 \lambda  (3 \mu  -1)+1}{8 \lambda }, &\quad \frac{1}{2-2 \mu -4 \nu }\leq \lambda \leq \frac{3}{6 \mu  -2} \\
			-\frac{4 \lambda ^2 \left(3 \mu ^2 -3 \mu  (1-\nu)+3 \nu ^2-3 \nu +1\right)+3}{6 \lambda }, &\quad \lambda >\frac{3}{6 \mu  -2} 
		\end{array}\right\},
		& \tfrac{1}{2}\leq\tfrac{\mu -\nu }{1-3 \nu }\leq\tfrac{2}{3}, 
		\\
		\left.
		\begin{array}{ll}
			-\mu +\nu,  &\quad 0\leq \lambda < \frac{1}{2 (2 \mu +\nu-1)} \\
			-\frac{4 \lambda  \left(\lambda  (2 \mu +\nu -1)^2-3 \nu +1\right)+1}{8 \lambda }, &\quad \frac{1}{2 \left(2 \mu +\nu -1\right)}\leq \lambda \leq \frac{3}{2 (1-3 \nu )}\\
			-\frac{4 \lambda ^2 \left(3 \mu ^2 -3 \mu  (1-\nu)+3 \nu ^2-3 \nu +1\right)+3}{6 \lambda }, &\quad \lambda >\frac{3}{2 (1-3 \nu )} 
		\end{array}\right\},
		& \tfrac{2}{3}\leq\tfrac{\mu -\nu }{1-3 \nu }\leq1,
	\end{array}
	\right.
	\label{GS_energy_inf}
\end{equation}
\end{widetext}
This expression is the generalization of the lowest-energy density introduced in \cite{nuestroPRE}, which is recovered by fixing $\nu=0$, where the remaining representations belong to U(2) subgroups of U(3) \cite{GREINER} and are known as the parental case. 

\begin{figure}[h]
	\begin{center}
		\includegraphics[
		width=1\columnwidth
		]{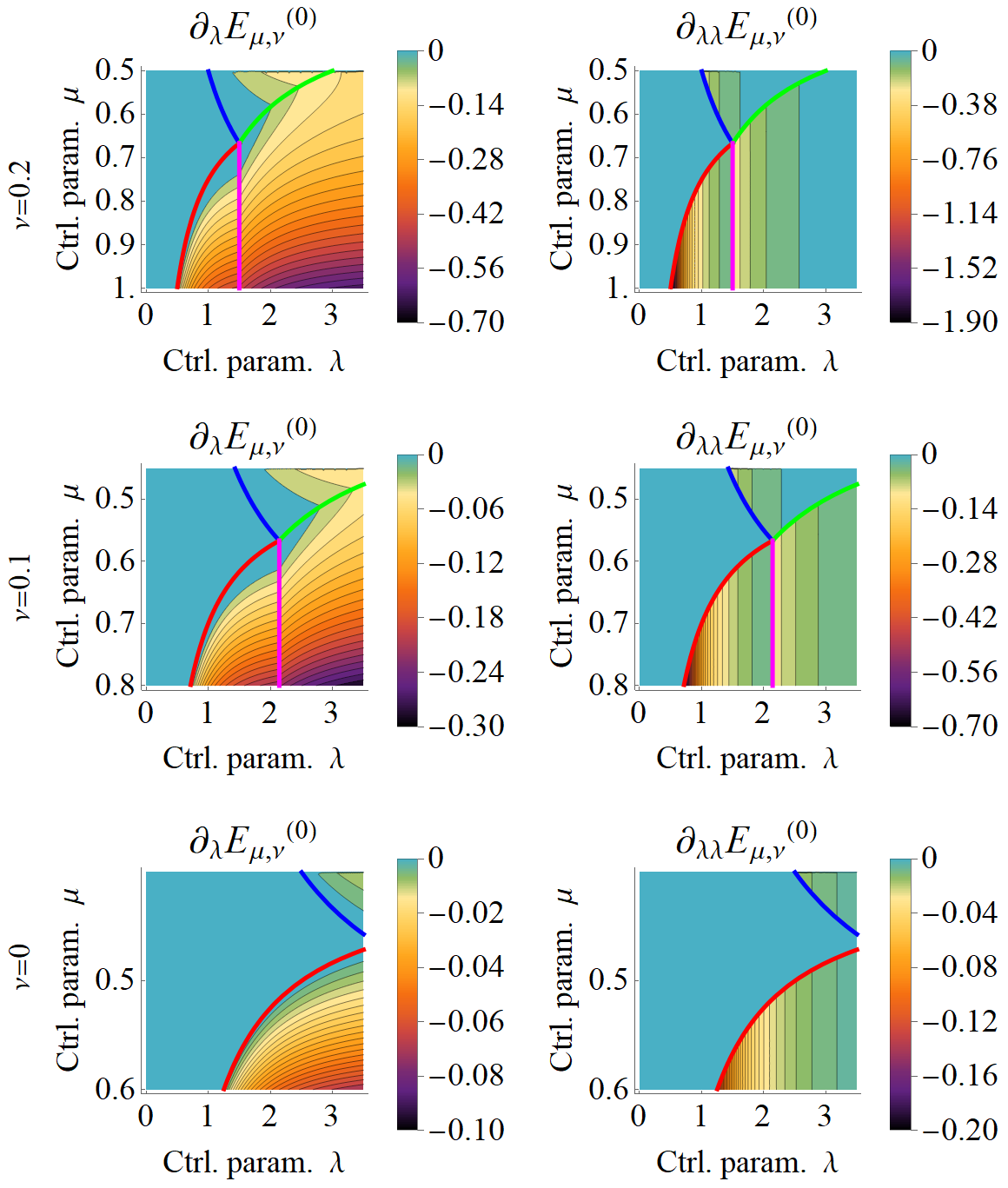}
	\end{center}
	\caption{Contour plots of the first (left panels) and second (right panels) $\lambda$-derivatives of the minimal energy density $E_{\mu,\nu}^{(0)}(\lambda)$  of the 3-level LMG model in the phase diagram $(\lambda,\mu)$ plane. Each column corresponds to a different value of the symmetry sector parameter $\nu\in[0,1/3]$. The vertical axis shrinks according to definition interval of $\mu$ \eqref{mu_nu_def} varying with $\nu$. The blue, green, red and magenta curves of critical points separate the four quantum  phases according to Eq. \eqref{mu_crit}.
	}
	\label{fig_Dlambda_Contour_Lambda-Mu}
\end{figure}

\begin{figure}[h]
	\begin{center}
		\includegraphics[
		width=1\columnwidth
		]{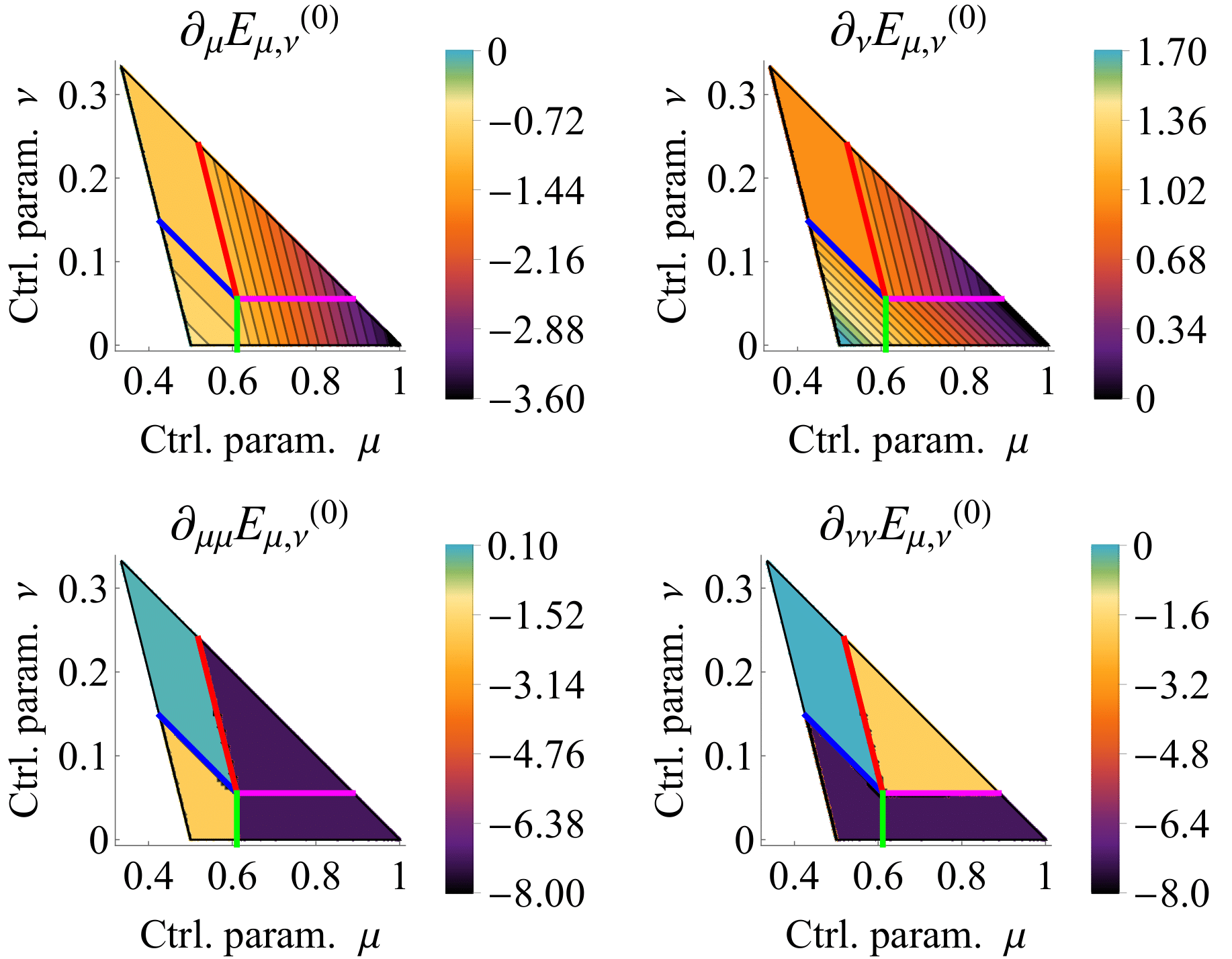}
	\end{center}
	\caption{Contour plots of the first (top panels) and second (bottom panels) $\mu$- and $\nu$-derivative of the lowest-energy density of the 3-level LMG model, in the phase diagram plane $(\mu,\nu)$ and at the interaction control parameter value $\lambda=1.8$. The limited triangular region of the plots is in accordance with the definition interval of $\mu,\nu$ \eqref{mu_nu_def}. The blue, green, red and magenta curves of critical points separate the model phases (see Eq.\eqref{nu_crit}). The top plots show how the Lieb-Mattis theorem is fulfilled,  $\partial_{\mu}E_{\mu,\nu}^{(0)}(\lambda)<0$ and $\partial_{\nu}E_{\mu,\nu}^{(0)}(\lambda)>0$ respectively for all $\mu,\nu$ (see Eq.\eqref{LM_theorem_derivatives}).}
	\label{fig_Dmu_Dnu_Contour_Mu-Nu}
\end{figure}

In the context of of MSQPT, the IR labels $\mu,\nu$ behave as two  additional control parameters, extending the $\lambda$-space phase diagram by $(\mu,\nu)$. Looking at the piecewise structure of \eqref{GS_energy_inf}, we see that the phase diagram presents four distinct quantum phases (denoted by: I, II, III and IV) in the $\lambda$-$\mu$-$\nu$ hyperplane. These four quantum phases coexist at a 
quadruple region defined by the curve $\{\lambda=\frac{3}{2-6\nu},\mu=\frac{1}{3}(2-3\nu)\}$, 
as it can be appreciated in Figures \ref{fig_Dlambda_Contour_Lambda-Mu} and \ref{fig_Dmu_Dnu_Contour_Mu-Nu}. The curves of critical points $\lambda_c$ separating two phases for a symmetry sector $(\mu,\nu)$ can be read  from the piecewise structure of the minimal energy surface \eqref{GS_energy_inf}, and have the following expressions
\begin{equation}
	\begin{array}{rcllll}
		\lambda_{c}^{\mathrm{I}\leftrightarrow\mathrm{IV}}(\mu,\nu)&=&\frac{1}{2-2\mu-4 \nu},&& \frac{1}{2}\leq \frac{\mu  -\nu }{1-3 \nu }\leq \frac{2}{3} &(\mathrm{blue})\\
		\lambda_{c}^{\mathrm{III}\leftrightarrow\mathrm{IV}}(\mu,\nu)&=&\frac{3}{6 \mu  -2},&& \frac{1}{2}\leq \frac{\mu  -\nu }{1-3 \nu }\leq \frac{2}{3} & (\mathrm{green})\\
		\lambda_{c}^{\mathrm{I}\leftrightarrow\mathrm{II}}(\mu,\nu)&=&\frac{1}{2 (2 \mu +\nu-1)}, &&\frac{2}{3}<\frac{\mu -\nu }{1-3 \nu }\leq 1&(\mathrm{red})\\
		\lambda_{c}^{\mathrm{II}\leftrightarrow\mathrm{III}}(\mu,\nu)&=&\frac{3}{2 (1-3 \nu )} ,&& \frac{2}{3}<\frac{\mu -\nu }{1-3 \nu }\leq 1 & (\mathrm{magenta})
	\end{array}\label{lambda_crit}
\end{equation}
It should be stressed that for $\nu=1/3=\mu$ (corresponding to the one-dimensional IR), these expressions of $\lambda_c$ diverge, and therefore there is no phase transition in this case. In some cases it is more convenient to express these curves as a function $\mu_c(\lambda,\nu)$, when plotting contour plots in the space $(\lambda,\mu)$, with $\nu$ as an hyperparameter, as in Figure \ref{fig_Dlambda_Contour_Lambda-Mu}. A similar transformation can be done to obtain the function $\nu_c(\mu,\lambda)$, when plotting in the space $(\mu,\nu)$ with $\lambda$ as hyperparameter, as in Figure \ref{fig_Dmu_Dnu_Contour_Mu-Nu}. The explicit expression of this functions can be found in Appendix \ref{appB}.

We show in Figure \ref{fig_Dlambda_Contour_Lambda-Mu} contour plots of the first (top panel) and second (bottom panel) $\lambda$-derivatives of the minimum energy $E_\mu^{(0)}(\epsilon,\lambda)$, in the $(\lambda,\mu)$ phase diagram, and using several values of $\nu$ in each column as hyperparameter. The extension to other values of $\nu$ can be seen in the animated GIF added in the supplementary material \cite{Supplemental}. From these contour plots we extract that the first derivatives of the energy density are continuous, while the second derivatives are discontinuous at
the critical curves \eqref{mu_crit}. This serves as proof that the MSQPT are second-order (in the Ehrenfest classification) for all values of the hyperparameter $\nu\in(0,1/3)$ in the graphics of Figure \ref{fig_Dlambda_Contour_Lambda-Mu}, as an  extension of the parental case $\nu=0$ \cite{nuestroPRE}.

Extending the  Lieb-Mattis theorem \cite{Lieb-Mattis_PR1962} to the thermodynamic limit, and labelling two representations $(\mu,\nu)$ and $(\mu',\nu')$ according to Eq. \eqref{mu_nu_def}, we can predict which of them would have the smaller  energy density $E_{\mu,\nu}^{(0)}$. In particular, according to equation \eqref{DominanceIneq}, the representation $(\mu',\nu')$ can be poured into $(\mu,\nu)$ if the inequalities $\mu\geq\mu'$ and $\nu\leq\nu'$ hold, leading to $E_{\mu,\nu}^{(0)}\leq E_{\mu',\nu'}^{(0)}$. This is equivalent to fulfill the expressions 
\begin{equation}
	\partial_\mu E_{\mu,\nu}^{(0)}(\lambda)\leq 0\,,\quad\quad \partial_\nu E_{\mu,\nu}^{(0)}(\lambda)\geq 0\,,\quad\quad \forall \lambda,\mu,\nu.\label{LM_theorem_derivatives}
\end{equation}
Both inequalities can be proved applying the derivative to each case in the piecewise expression of the lowest-energy density \eqref{GS_energy_inf}. For the first $\mu$-derivative we have
\begin{equation}
	\partial_\mu E_{\mu,\nu}^{(0)}(\lambda)=\left\{ 
	\begin{array}{ll}
		-1,  & \text{Cases 1,4}, \\
		-\frac{3}{2}-\lambda(\mu+2\nu-1), &\text{Case 2}, \\
		-2\lambda(2\mu+\nu-1), &\text{Cases 3,5,6} .
	\end{array}\right.\label{GS_Dmu}
\end{equation}
The first and fourth cases trivially give $\partial_\mu E_{\mu,\nu}^{(0)}(\lambda)=-1<0$. The third, fifth and sixth cases are negative  from the inequalities $\lambda>0$ and $\mu>\tfrac{1}{2}(1-\nu)$ by definition \eqref{mu_nu_def}. 
The second case can be proved using Reductio ad absurdum. Supose that $-\tfrac{3}{2}-\lambda(\mu+2\nu-1)>0$, hence $\lambda> \tfrac{3}{2(1-\mu-2\nu)}$. At the same time, we can transform the definition interval $\mu\geq \frac{1}{2}(1-\nu)$ into the following inequality  $\tfrac{3}{2(1-\mu-2\nu)}\geq \tfrac{3}{6\mu-2}$. Merging this expression with the previous one $\lambda> \tfrac{3}{2(1-\mu-2\nu)}$, we arrive to $\lambda>\tfrac{3}{6\mu-2}$, which is out of the case 2 in equation \eqref{GS_Dmu}, concluding the proof.

The analysis of the positivity of the first $\nu$-derivative 
\begin{equation}
	\partial_\nu E_{\mu,\nu}^{(0)}(\lambda)=
	\left\{ 
	\begin{array}{ll}
		-1,  & \text{Cases 1,4}, \\
		-2\lambda(\mu+2\nu-1), &\text{Cases 2,3,6}, \\
		\frac{3}{2}-\lambda(2\mu+\nu-1), &\text{Case 5},
	\end{array}\right.
	\label{GS_Dnu}
\end{equation}
follows steps analogous to those of the $\mu$-derivative, except that now we use that $\mu\leq 1-2\nu$ according to \eqref{mu_nu_def}. The application of the Lieb-Mattis theorem to the LMG model \eqref{LM_theorem_derivatives} can also be graphically represented  like in Figure \ref{fig_Dmu_Dnu_Contour_Mu-Nu}, where we show the $\mu-$ and $\nu$-derivatives of the energy density $E_{\mu,\nu}^{(0)}(\lambda)$ in the phase diagram plane $(\mu,\nu)$ for the particular case $\lambda=1.8$. The extension to other values of $\lambda$ can be seen in the animated GIF added in the supplementary material \cite{Supplemental}. In the first contour plot of Figure \ref{fig_Dmu_Dnu_Contour_Mu-Nu}, the $\mu$-derivative of the energy density is negative in all the allowed $(\mu,\nu)$ region, while the second contour plot shows the positivity  of the $\nu$-derivative. The restriction to the triangular region is due to the definition interval of $\mu,\nu$ in \eqref{mu_nu_def}. The third and fourth contour plots display how the second derivatives are discontinuous at the critical curves \eqref{nu_crit}, thus underlining the continuous (second-order) character  of the MSQPT. Despite the second $\mu$- and $\nu$-derivative are continuous in the magenta and green lines phase transitions respectively, the characterization of second-order MSQPT holds in general.

\begin{figure}[h]
	\begin{center}
		\includegraphics[
		width=1\columnwidth
		]{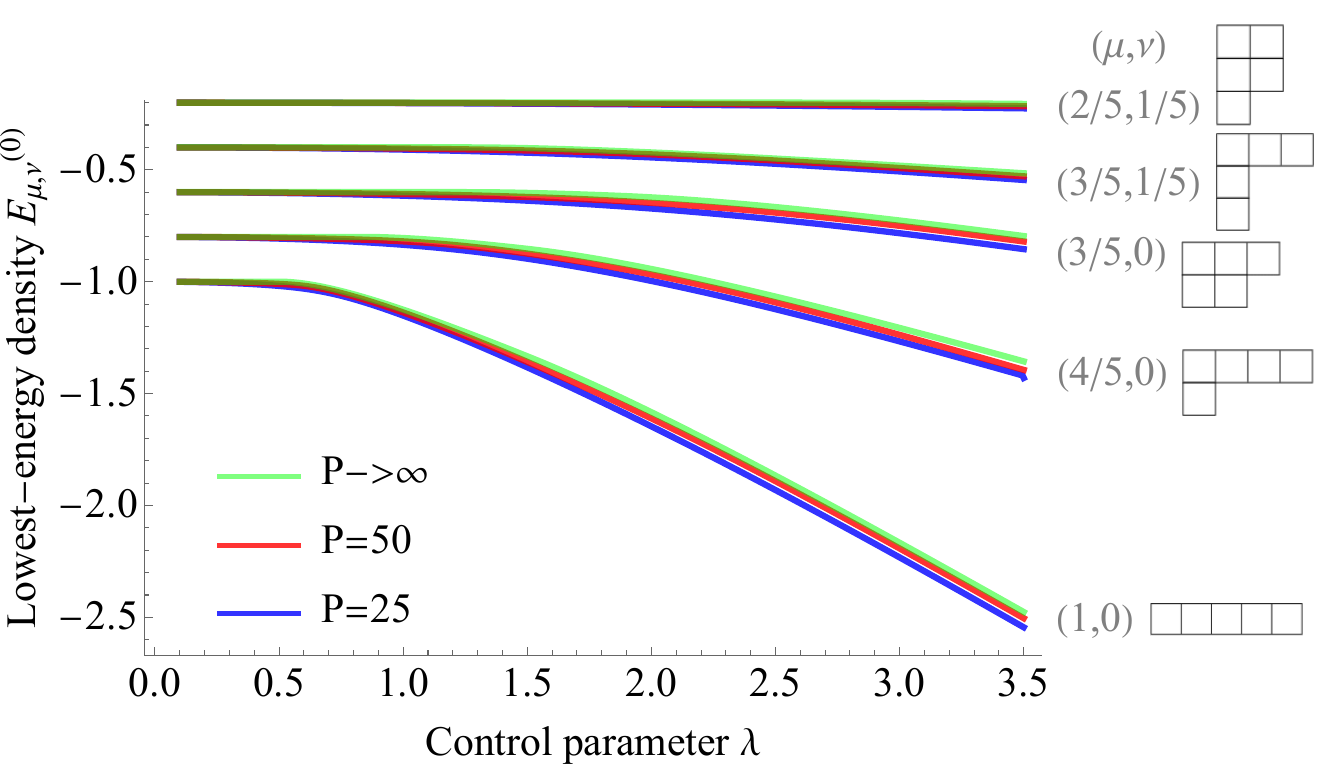}
	\end{center}
	\caption{Lowest-energy density of the 3-level LMG Hamiltonian  as a function of the control parameter $\lambda$, for five different U(3) IRs labelled by $(\mu,\nu)$ \eqref{mu_nu_def}. The energy densities are numerically computed for a finite number of $P=25$ and $P=50$ particles, and in the thermodynamic limit $P\to\infty$ using the analytic expression \eqref{GS_energy_inf}. At the right hand side of every representation label $(\mu,\nu)$, we show the equivalent Young diagram for $P=5$ particles \eqref{5qutrits}. The Young diagrams for $P=25$ and $P=50$ could be obtained multiplying by 5 and 10 times, respectively, the diagrams shown for $P=5$.}
	\label{fig1}
\end{figure}

The analytical results in the thermodynamic limit $P\to\infty$ can be complemented and compared with numerical results for the finite $P$ case, where the LMG Hamiltonian \eqref{Hd} is a square matrix of size $3^P$ to diagonalize. As the interaction does not mix different symmetry sectors, the Hamiltonian adopts  a block diagonal form when the basis vectors are adapted to IRs $h$ of the Lie group U($N$). Hence, the collective operators $S_{ij}$ generating the group U($N$) are computed in the Gelfand-Tsetlin basis \cite{Barut}, where their particular matrix expression depends on the chosen representation \cite{nuestroPRE}. For example, using the Young diagrammatic method \cite{CVI,Barut,Thrall1954,HallBook,ZHAO},  the tensor product space \eqref{tensorprod} of $P=5$ qutrits ($N=3$, $M=1$, $L=5$), can be decomposed into IRs as 

\begin{multline}
\Yvcentermath1 \stackrel{3}{\yng(1)}\otimes \stackrel{3}{\yng(1)}\otimes \Yvcentermath1\stackrel{3}{\yng(1)} \otimes \stackrel{3}{\yng(1)}
\otimes \stackrel{3}{\yng(1)}
\:\: \\
=  \Yvcentermath1\:\:\stackrel{21}{\yng(5)}\:\oplus\: 4\stackrel{24}{\yng(4,1)}\:\oplus\: 5\stackrel{15}{\yng(3,2)}\:\\
\Yvcentermath1 \oplus\: 6\stackrel{6}{\yng(3,1,1)}\:\oplus\: 5\stackrel{3}{\yng(2,2,1)}\:\:, \label{5qutrits} 
\end{multline}
or, equivalently, in the standard notation  $[h_1,h_2,h_3]$ of Eq. \eqref{youngdiagram}
\begin{equation}
 [1]^{\otimes 5} = [5,0,0]\oplus 4[4,1,0] \oplus 5[3,2,0]\oplus 6[3,1,1]\oplus 5[2,2,1],
\end{equation}
where we are placing the corresponding IR dimensions \eqref{dimh} on top of each Young diagram, so that the equality $3^5=21+4*24+5*15+6*6+5*3$ is verified. The notation  $[h_1,h_2,h_3]$ can be transformed into the $(\mu,\nu)$ notation by simply using the expression \eqref{mu_nu_def}. This assignations are shown in the right side of Figure \ref{fig1}. The $(\mu,\nu)$ notation becomes useful when comparing two IRs $h$ and $h'$  of the same proportions  $(\mu,\nu)$ but a different number of particles $P$. For instance, in Figure \ref{fig1}, we see the ground state energy of the 3-level LMG Hamiltonian as a function of $\lambda$ in the three ($P=25, 50, \infty$) bottom-most cluster of curves (blue, red and green), corresponding to the totally symmetric IR $(1,0)$. The energy density increases with $P$ and tends to $E_{\mu,\nu}^{(0)}(\lambda)$ (green curve) in the thermodynamic limit as expected, thus proving the convergence of the numeric to the analytic results \eqref{GS_energy_inf}. The comparison would not have been possible without using the $(\mu,\nu)$ notation as intensive parameter for the representations. In addition, for all considered  IRs $(\mu,\nu)$, the Lieb-Mattis theorem holds, as expected, where the higher is $\mu$ and the lower is $\nu$ (moving downwards in Figure \ref{fig1}), the lower is the minimum energy density $E_{\mu,\nu}^{(0)}$, in agreement with \eqref{LM_theorem_derivatives}. The data that support the findings of this article are openly available \cite{DiagonalizationData}.

\section{Conclusions and outlook}\label{conclusec}

The subject of quantum phase transitions is enriched when we increase the number $N$ of internal components of the essential constituents or particles, and when we incorporate to the study, not only the ground state of the system belonging to the most symmetric sector, but all symmetry sectors $h$ with mixed symmetries (fermionic mixtures), arising in the $P$-fold tensor product Clebsch-Gordan decomposition for $P$ particles, in the thermodynamic limit $P\to\infty$. Mixed symmetry sectors $h$ refer to both: permutation symmetry $S_N$ and the different irreducible representations of the underlying symmetry group $\rmu(N)$ \cite{Barut}. Since the Hamiltonian is $\rmu(N)$-invariant, temporal evolution does not mix symmetry sectors $h$. Therefore, QPTs can occur inside each mixed  symmetry sector $h$ for critical values $\lambda_c(h)$ of the interaction strength parameter $\lambda$, thus extending the concept of QPT to Mixed Symmetry QPT (MSQPT) and the phase diagram can be enlarged from $\lambda$ to $(\lambda,h)$.  The Lieb-Mattis theorem provides an ordering of the lowest-energy Hamiltonian eigenstates inside each symmetry sector $h$ through dominance relations. We show how  this ordering can be extended to the thermodynamic limit through monotonic properties of energy density functions in the extended phase diagram  $(\lambda,h)$. Coherent states  on  Grassmannian phase spaces  $\rmu(N)/\rmu(M)\times \rmu(N-M)$ (for filling factor $M$) turn out to be good variational quasi-classical states reproducing the ground state system inside each mixed permutation symmetry sector, specially in the thermodynamic limit.

We have exemplified this general construction with a generalization of the traditional  Lipkin-Meshkov-Glick Hamiltonian model for $P$ identical two-level particles to $N$ levels, focusing on the cases $N=2$ and $N=3$. The LMG model undergoes a second order QPT for certain critical  values $\lambda_c$ of the interaction parameter $\lambda$.  These critical  $\lambda_c$ values are affected when we move from the ground state (fully symmetric) sector $h_0\sim (\mu=1,\nu=0)$ to other permutation symmetry sectors $h\sim(\mu,\nu)$. We show that the corresponding lowest energy density states inside each sector $(\mu,\nu)$ are still ordered according to the Lieb-Mattis theorem, even for $P\to\infty$, where dominance ordering is replaced by analytic and monotonic properties of the energy density $E_{\mu,\nu}^{(0)}(\lambda)$ with respect to the \emph{continuous} variables $(\mu,\nu)$. The phase diagram of the system is enriched when we introduce the continuous symmetry sector labels $(\mu,\nu)$ as extra control parameters in addition to the interaction strength $\lambda$, with the appearance of new quantum phases as regards standard  QPTs occurring in the fully symmetric sector $(\mu=1,\nu=0)$.

\section*{Acknowledgments}
We thank the support of the Spanish Ministry of Science through the project PID2022-138144NB-I00. 

%
%
%

\appendix
\section{LMG energy surface and minimization for $N=3$ level particles}\label{app}

We calculate the energy surface of the 3-level LMG model in the thermodynamic limit using the Hamiltonian \eqref{Hd}, the CS expectation values for U(3) in \eqref{CSEV}, and the IRs parametrization of the eq.\eqref{mu_nu_def}.
\begin{widetext}
\begin{align}
	E^{Z}_{\mu,\nu}(\lambda)=
		\,\tfrac{1}{|Z^{\dagger}Z|_1 |Z^{\dagger}Z|_2}&\Bigg[
		(z_1+z_2\bar{z}_3)(\bar{z}_1+\bar{z}_2 z_3)(\mu+\nu-1) 
		+|Z^{\dagger}Z|_1 \Big[\mu+\Big(2-(z_2-z_1 z_3)(\bar{z}_2-\bar{z}_1 \bar{z}_3)\Big)\nu-1\Big]\nonumber\\
		&	-|Z^{\dagger}Z|_2 \Big[ (2-\bar{z}_2 z_2)\mu + \nu -1 + \bar{z}_1 z_1(\mu+\nu-1) \Big]
		\Bigg]\nonumber\\
		\,-\tfrac{\lambda}{|Z^{\dagger}Z|_1^2 |Z^{\dagger}Z|_2^2}&\Bigg[
			|Z^{\dagger}Z|_2^2\Big[(z_1^2+(1+\bar{z}_1^2)z_2^2)(2\mu+\nu-1)^2\Big]\nonumber\\
		&	-2 |Z^{\dagger}Z|_1 |Z^{\dagger}Z|_2\Big[\Big((z_2+(\bar{z}_1-z_1)z_3)z_2 + (z_2-z_1 z_3)z_1 \bar{z}_3\Big)*(2\mu+\nu-1)(\mu+2\nu-1)\Big]\nonumber\\
		&+|Z^{\dagger}Z|_1^2\Big[\Big((z_2-2z_1 z_3)z_2 + (1+z_1^2)z_3^2+(z_2-z_1 z_3)^2 \bar{z}_3^2\Big)(\mu+2\nu-1)^2\Big]
		\Bigg].
\end{align}

The minimization of $E^{Z}_{\mu,\nu}(\lambda)$ in the CS complex coordinates $Z$ is performed using the polar form $z_j=|z_j|\exp(i\theta_j)$ for all $j=1,2,3$, and calculating the derivatives respect to the modulus $|z_j|$ and the argument $\theta_j$. The final solution is composed by real minima, where $\theta_j=0,\pi$ for all $j=1,2,3$, so that


\begin{equation}
	z_1^{(0)}(\lambda,\mu,\nu)=\pm\left\{  
	\begin{array}{ll}
		\left. 
		\begin{array}{ll}
			0,  &\quad 0\leq \lambda < \frac{3}{6 \mu  -2} \\
			2\sqrt{\frac{\lambda(1-3\nu)(2-3\mu-3\nu)}{(1-3\mu)(3+2\lambda(1-3\nu))}}, &\quad \lambda \geq\frac{3}{6 \mu  -2} 
		\end{array}\right\},
		& \text{Cond.1}, 
		\\
		\left.
		\begin{array}{ll}
			0,  &\quad 0\leq \lambda < \frac{1}{2 (2 \mu +\nu-1)} \\
			\sqrt{\frac{-1+2\lambda(2\mu+\nu-1)}{1+2\lambda(2\mu+\nu-1)}}, &\quad \frac{1}{2 \left(2 \mu +\nu -1\right)}\leq \lambda \leq \frac{3}{2 (1-3 \nu )}\\
			2\sqrt{\frac{\lambda(1-3\nu)(2-3\mu-3\nu)}{(1-3\mu)(3+2\lambda(1-3\nu))}}, &\quad \lambda >\frac{3}{2 (1-3 \nu )} 
		\end{array}\right\},
		& \text{Cond.2},
	\end{array}
	\right.
	\label{alpha0}
\end{equation}


\begin{equation}
	z_2^{(0)}(\lambda,\mu,\nu)=\pm\left\{  
	\begin{array}{ll}
		\left. 
		\begin{array}{ll}
			0,  &\quad 0\leq \lambda < \frac{3}{6 \mu  -2} \\
			\sqrt{1-\frac{6}{3+2\lambda(1-3\nu)}}, &\quad \lambda \geq \frac{3}{6 \mu  -2} 
		\end{array}\right\},
		& \text{Cond.1}, 
		\\
		\left.
		\begin{array}{ll}
			0,  &\quad 0\leq \lambda < \frac{3}{2 (1-3 \nu )} \\
			\sqrt{1-\frac{6}{3+2\lambda(1-3\nu)}}, &\quad \lambda \geq\frac{3}{2 (1-3 \nu )} 
		\end{array}\right\},
		& \text{Cond.2},
	\end{array}
	\right.
	\label{beta0}
\end{equation}


\begin{equation}
	z_3^{(0)}(\lambda,\mu,\nu)=\pm\left\{  
	\begin{array}{ll}
		\left. 
		\begin{array}{ll}
			0,  &\quad 0\leq \lambda < \frac{1}{2-2\mu-4 \nu } \\
			\sqrt{\frac{1+2\lambda(\mu+2\nu-1)}{-1+2\lambda(\mu+2\nu-1)}}, &\quad \frac{1}{2-2 \mu -4 \nu }\leq \lambda \leq \frac{3}{6 \mu  -2} \\
			0, &\quad \lambda >\frac{3}{6 \mu  -2} 
		\end{array}\right\},
		& \text{Cond.1}, 
		\\
		\left.
		\begin{array}{ll}
			0,  &\quad \lambda \geq 0  
		\end{array}\right.,
		& \text{Cond.2},
	\end{array}
	\right.
	\label{gamma0}
\end{equation}
where the condition Cond.1 stands for $\tfrac{1}{2}\leq\tfrac{\mu -\nu }{1-3 \nu }\leq\tfrac{2}{3}$ and Cond.2 for $\tfrac{2}{3}\leq\tfrac{\mu -\nu }{1-3 \nu }\leq1$. The minimum energy surface expression is given in the main text \eqref{GS_energy_inf}.

\end{widetext}

\section{Curves of critical points for $N=3$ level LMG}\label{appB}

According to the structure of the minimal energy surface \eqref{GS_energy_inf}, the phase diagram presents four distinct quantum phases in the $\lambda$-$\mu$-$\nu$ hyperplane, as depicted in Figures \ref{fig_Dlambda_Contour_Lambda-Mu} and \ref{fig_Dmu_Dnu_Contour_Mu-Nu}. In equation \eqref{lambda_crit}, we represented the curves of critical points $\lambda_c$ separating combinations of two phases, as functions of the symmetry sector parameters $(\mu,\nu)$. In order to fully characterize the MSQPT order, it is more convenient the compute critical curves as functions in the $(\lambda,\nu)$ subspace,

{\small
\begin{equation}
	\begin{array}{rcllll}
		\mu_{c}^{\mathrm{I}\leftrightarrow\mathrm{IV}}(\lambda,\nu)&=&\frac{2\lambda(1-2\nu)-1}{2\lambda},&& 1\leq \lambda(1-3\nu)\leq \frac{3}{2} &(\mathrm{b})\\[1ex]
		\mu_{c}^{\mathrm{III}\leftrightarrow\mathrm{IV}}(\lambda,\nu)&=&\frac{3+2\lambda}{6\lambda},&& \frac{3}{2}<\lambda(1-3\nu)\leq 3 & (\mathrm{g})\\[1ex]
		\mu_{c}^{\mathrm{I}\leftrightarrow\mathrm{II}}(\lambda,\nu)&=&\frac{2\lambda(1-\nu)+1}{4\lambda}, && \frac{1}{2}<\lambda(1-3\nu)\leq \frac{3}{2} &(\mathrm{r})\\[1ex]
		\mu_{c}^{\mathrm{II}\leftrightarrow\mathrm{III}}(\lambda,\nu)&\in&\Big(\frac{1}{3}{\scriptstyle(2-3\nu)},{\scriptstyle 1-2\nu}\Big) ,&& \lambda=\frac{3}{2(1-3\nu)} & (\mathrm{m})
	\end{array}\label{mu_crit}
\end{equation}
}

That is the case of Figure \ref{fig_Dlambda_Contour_Lambda-Mu}. In contrast, to enlighten the Lieb-Mattis theorem \eqref{DominanceIneq} in Figure \ref{fig_Dmu_Dnu_Contour_Mu-Nu}, we must express the critical curves in the $(\mu,\lambda)$ subspace,

{\small
\begin{equation}
	\begin{array}{rcllll}
		\nu_{c}^{\mathrm{I}\leftrightarrow\mathrm{IV}}(\mu,\lambda)&=&\frac{2\lambda(1-\mu)-1}{4\lambda},&& \frac{1}{2}\leq \lambda(3\mu-1)\leq \frac{3}{2} &(\mathrm{b})\\[1ex]
		\nu_{c}^{\mathrm{III}\leftrightarrow\mathrm{IV}}(\mu,\lambda)&\in&(\frac{\lambda-3}{3\lambda},\frac{2\lambda-3}{6\lambda}),&& \mu=\frac{2\lambda+3}{6\lambda} & (\mathrm{g})\\[1ex]
		\nu_{c}^{\mathrm{I}\leftrightarrow\mathrm{II}}(\mu,\lambda)&=&\frac{2\lambda(1-2\mu)+1}{2\lambda}, && 1\leq \lambda(3\mu-1)\leq \frac{3}{2} &(\mathrm{r})\\[1ex]
		\nu_{c}^{\mathrm{II}\leftrightarrow\mathrm{III}}(\mu,\lambda)&=& \frac{-3+2\lambda}{6\lambda} ,&& \frac{3}{2}\leq \lambda(3\mu-1)\leq 3 & (\mathrm{m})
	\end{array}\label{nu_crit}
\end{equation}
}
The labels at the right side of the equations stand for the line colors blue (b), green (g), red (r) and magenta (m) in Figures \ref{fig_Dlambda_Contour_Lambda-Mu} and \ref{fig_Dmu_Dnu_Contour_Mu-Nu}.

\bibliography{bibliografia.bib}

\begin{thebibliography}{59}%
\makeatletter
\providecommand \@ifxundefined [1]{%
 \@ifx{#1\undefined}
}%
\providecommand \@ifnum [1]{%
 \ifnum #1\expandafter \@firstoftwo
 \else \expandafter \@secondoftwo
 \fi
}%
\providecommand \@ifx [1]{%
 \ifx #1\expandafter \@firstoftwo
 \else \expandafter \@secondoftwo
 \fi
}%
\providecommand \natexlab [1]{#1}%
\providecommand \enquote  [1]{``#1''}%
\providecommand \bibnamefont  [1]{#1}%
\providecommand \bibfnamefont [1]{#1}%
\providecommand \citenamefont [1]{#1}%
\providecommand \href@noop [0]{\@secondoftwo}%
\providecommand \href [0]{\begingroup \@sanitize@url \@href}%
\providecommand \@href[1]{\@@startlink{#1}\@@href}%
\providecommand \@@href[1]{\endgroup#1\@@endlink}%
\providecommand \@sanitize@url [0]{\catcode `\\12\catcode `\$12\catcode
  `\&12\catcode `\#12\catcode `\^12\catcode `\_12\catcode `\%12\relax}%
\providecommand \@@startlink[1]{}%
\providecommand \@@endlink[0]{}%
\providecommand \url  [0]{\begingroup\@sanitize@url \@url }%
\providecommand \@url [1]{\endgroup\@href {#1}{\urlprefix }}%
\providecommand \urlprefix  [0]{URL }%
\providecommand \Eprint [0]{\href }%
\providecommand \doibase [0]{https://doi.org/}%
\providecommand \selectlanguage [0]{\@gobble}%
\providecommand \bibinfo  [0]{\@secondoftwo}%
\providecommand \bibfield  [0]{\@secondoftwo}%
\providecommand \translation [1]{[#1]}%
\providecommand \BibitemOpen [0]{}%
\providecommand \bibitemStop [0]{}%
\providecommand \bibitemNoStop [0]{.\EOS\space}%
\providecommand \EOS [0]{\spacefactor3000\relax}%
\providecommand \BibitemShut  [1]{\csname bibitem#1\endcsname}%
\let\auto@bib@innerbib\@empty
\bibitem [{\citenamefont {Decamp}\ \emph {et~al.}(2016)\citenamefont {Decamp},
  \citenamefont {Armagnat}, \citenamefont {Fang}, \citenamefont {Albert},
  \citenamefont {Minguzzi},\ and\ \citenamefont {Vignolo}}]{Decamp16}%
  \BibitemOpen
  \bibfield  {author} {\bibinfo {author} {\bibfnamefont {J.}~\bibnamefont
  {Decamp}}, \bibinfo {author} {\bibfnamefont {P.}~\bibnamefont {Armagnat}},
  \bibinfo {author} {\bibfnamefont {B.}~\bibnamefont {Fang}}, \bibinfo {author}
  {\bibfnamefont {M.}~\bibnamefont {Albert}}, \bibinfo {author} {\bibfnamefont
  {A.}~\bibnamefont {Minguzzi}},\ and\ \bibinfo {author} {\bibfnamefont
  {P.}~\bibnamefont {Vignolo}},\ }\bibfield  {title} {\bibinfo {title} {Exact
  density profiles and symmetry classification for strongly interacting
  multi-component fermi gases in tight waveguides},\ }\href
  {https://doi.org/10.1088/1367-2630/18/5/055011} {\bibfield  {journal}
  {\bibinfo  {journal} {New J. Phys.}\ }\textbf {\bibinfo {volume} {18}},\
  \bibinfo {pages} {055011} (\bibinfo {year} {2016})}\BibitemShut {NoStop}%
\bibitem [{\citenamefont {Decamp}\ \emph
  {et~al.}(2020{\natexlab{a}})\citenamefont {Decamp}, \citenamefont {Gong},
  \citenamefont {Loh},\ and\ \citenamefont
  {Miniatura}}]{PhysRevResearch.2.023059}%
  \BibitemOpen
  \bibfield  {author} {\bibinfo {author} {\bibfnamefont {J.}~\bibnamefont
  {Decamp}}, \bibinfo {author} {\bibfnamefont {J.}~\bibnamefont {Gong}},
  \bibinfo {author} {\bibfnamefont {H.}~\bibnamefont {Loh}},\ and\ \bibinfo
  {author} {\bibfnamefont {C.}~\bibnamefont {Miniatura}},\ }\bibfield  {title}
  {\bibinfo {title} {Graph-theory treatment of one-dimensional strongly
  repulsive fermions},\ }\href
  {https://doi.org/10.1103/PhysRevResearch.2.023059} {\bibfield  {journal}
  {\bibinfo  {journal} {Phys. Rev. Res.}\ }\textbf {\bibinfo {volume} {2}},\
  \bibinfo {pages} {023059} (\bibinfo {year} {2020}{\natexlab{a}})}\BibitemShut
  {NoStop}%
\bibitem [{\citenamefont {Pethick}\ and\ \citenamefont
  {Smith}(2008)}]{pethick_smith_2008}%
  \BibitemOpen
  \bibfield  {author} {\bibinfo {author} {\bibfnamefont {C.~J.}\ \bibnamefont
  {Pethick}}\ and\ \bibinfo {author} {\bibfnamefont {H.}~\bibnamefont
  {Smith}},\ }\href {https://doi.org/10.1017/CBO9780511802850} {\emph {\bibinfo
  {title} {Bose-Einstein Condensation in Dilute Gases}}},\ \bibinfo {edition}
  {2nd}\ ed.\ (\bibinfo  {publisher} {Cambridge University Press},\ \bibinfo
  {year} {2008})\BibitemShut {NoStop}%
\bibitem [{\citenamefont {Lewenstein}\ \emph {et~al.}(2012)\citenamefont
  {Lewenstein}, \citenamefont {Sanpera},\ and\ \citenamefont
  {Ahufinger}}]{Lewenstein_2012}%
  \BibitemOpen
  \bibfield  {author} {\bibinfo {author} {\bibfnamefont {M.}~\bibnamefont
  {Lewenstein}}, \bibinfo {author} {\bibfnamefont {A.}~\bibnamefont
  {Sanpera}},\ and\ \bibinfo {author} {\bibfnamefont {V.}~\bibnamefont
  {Ahufinger}},\ }\href
  {https://doi.org/10.1093/acprof:oso/9780199573127.001.0001} {\emph {\bibinfo
  {title} {Ultracold Atoms in Optical Lattices: Simulating quantum many-body
  systems}}},\ \bibinfo {edition} {1st}\ ed.\ (\bibinfo  {publisher} {Oxford
  University Press},\ \bibinfo {year} {2012})\BibitemShut {NoStop}%
\bibitem [{\citenamefont {Cazalilla}\ and\ \citenamefont
  {Rey}(2014)}]{Cazalilla_2014}%
  \BibitemOpen
  \bibfield  {author} {\bibinfo {author} {\bibfnamefont {M.~A.}\ \bibnamefont
  {Cazalilla}}\ and\ \bibinfo {author} {\bibfnamefont {A.~M.}\ \bibnamefont
  {Rey}},\ }\bibfield  {title} {\bibinfo {title} {Ultracold fermi gases with
  emergent {SU}(n) symmetry},\ }\href
  {https://doi.org/10.1088/0034-4885/77/12/124401} {\bibfield  {journal}
  {\bibinfo  {journal} {Reports on Progress in Physics}\ }\textbf {\bibinfo
  {volume} {77}},\ \bibinfo {pages} {124401} (\bibinfo {year}
  {2014})}\BibitemShut {NoStop}%
\bibitem [{\citenamefont {Sowi{\'{n}}ski}\ and\ \citenamefont
  {Garc{\'{\i}}a-March}(2019)}]{Sowinski19}%
  \BibitemOpen
  \bibfield  {author} {\bibinfo {author} {\bibfnamefont {T.}~\bibnamefont
  {Sowi{\'{n}}ski}}\ and\ \bibinfo {author} {\bibfnamefont {M.~{\'{A}}.}\
  \bibnamefont {Garc{\'{\i}}a-March}},\ }\bibfield  {title} {\bibinfo {title}
  {One-dimensional mixtures of several ultracold atoms: a review},\ }\href
  {https://doi.org/10.1088/1361-6633/ab3a80} {\bibfield  {journal} {\bibinfo
  {journal} {Rep. Prog. Phys.}\ }\textbf {\bibinfo {volume} {82}},\ \bibinfo
  {pages} {104401} (\bibinfo {year} {2019})}\BibitemShut {NoStop}%
\bibitem [{\citenamefont {Zhang}\ \emph {et~al.}(2014)\citenamefont {Zhang},
  \citenamefont {Bishof}, \citenamefont {Bromley}, \citenamefont {Kraus},
  \citenamefont {Safronova}, \citenamefont {Zoller}, \citenamefont {Rey},\ and\
  \citenamefont {Ye}}]{doi:10.1126/science.1254978}%
  \BibitemOpen
  \bibfield  {author} {\bibinfo {author} {\bibfnamefont {X.}~\bibnamefont
  {Zhang}}, \bibinfo {author} {\bibfnamefont {M.}~\bibnamefont {Bishof}},
  \bibinfo {author} {\bibfnamefont {S.~L.}\ \bibnamefont {Bromley}}, \bibinfo
  {author} {\bibfnamefont {C.~V.}\ \bibnamefont {Kraus}}, \bibinfo {author}
  {\bibfnamefont {M.~S.}\ \bibnamefont {Safronova}}, \bibinfo {author}
  {\bibfnamefont {P.}~\bibnamefont {Zoller}}, \bibinfo {author} {\bibfnamefont
  {A.~M.}\ \bibnamefont {Rey}},\ and\ \bibinfo {author} {\bibfnamefont
  {J.}~\bibnamefont {Ye}},\ }\bibfield  {title} {\bibinfo {title}
  {Spectroscopic observation of su(<i>n</i>)-symmetric interactions in sr
  orbital magnetism},\ }\href {https://doi.org/10.1126/science.1254978}
  {\bibfield  {journal} {\bibinfo  {journal} {Science}\ }\textbf {\bibinfo
  {volume} {345}},\ \bibinfo {pages} {1467} (\bibinfo {year} {2014})},\ \Eprint
  {https://arxiv.org/abs/https://www.science.org/doi/pdf/10.1126/science.1254978}
  {https://www.science.org/doi/pdf/10.1126/science.1254978} \BibitemShut
  {NoStop}%
\bibitem [{\citenamefont {Honerkamp}\ and\ \citenamefont
  {Hofstetter}(2004)}]{PhysRevLett.92.170403}%
  \BibitemOpen
  \bibfield  {author} {\bibinfo {author} {\bibfnamefont {C.}~\bibnamefont
  {Honerkamp}}\ and\ \bibinfo {author} {\bibfnamefont {W.}~\bibnamefont
  {Hofstetter}},\ }\bibfield  {title} {\bibinfo {title} {Ultracold fermions and
  the $\mathrm{SU}(n)$ hubbard model},\ }\href
  {https://doi.org/10.1103/PhysRevLett.92.170403} {\bibfield  {journal}
  {\bibinfo  {journal} {Phys. Rev. Lett.}\ }\textbf {\bibinfo {volume} {92}},\
  \bibinfo {pages} {170403} (\bibinfo {year} {2004})}\BibitemShut {NoStop}%
\bibitem [{\citenamefont {Zhang}\ \emph {et~al.}(2019)\citenamefont {Zhang},
  \citenamefont {Vidmar},\ and\ \citenamefont {Rigol}}]{PhysRevA.99.063605}%
  \BibitemOpen
  \bibfield  {author} {\bibinfo {author} {\bibfnamefont {Y.}~\bibnamefont
  {Zhang}}, \bibinfo {author} {\bibfnamefont {L.}~\bibnamefont {Vidmar}},\ and\
  \bibinfo {author} {\bibfnamefont {M.}~\bibnamefont {Rigol}},\ }\bibfield
  {title} {\bibinfo {title} {Quantum dynamics of impenetrable $\mathrm{SU}(n)$
  fermions in one-dimensional lattices},\ }\href
  {https://doi.org/10.1103/PhysRevA.99.063605} {\bibfield  {journal} {\bibinfo
  {journal} {Phys. Rev. A}\ }\textbf {\bibinfo {volume} {99}},\ \bibinfo
  {pages} {063605} (\bibinfo {year} {2019})}\BibitemShut {NoStop}%
\bibitem [{\citenamefont {Cazalilla}\ \emph {et~al.}(2009)\citenamefont
  {Cazalilla}, \citenamefont {Ho},\ and\ \citenamefont
  {Ueda}}]{Cazalilla_2009}%
  \BibitemOpen
  \bibfield  {author} {\bibinfo {author} {\bibfnamefont {M.~A.}\ \bibnamefont
  {Cazalilla}}, \bibinfo {author} {\bibfnamefont {A.~F.}\ \bibnamefont {Ho}},\
  and\ \bibinfo {author} {\bibfnamefont {M.}~\bibnamefont {Ueda}},\ }\bibfield
  {title} {\bibinfo {title} {Ultracold gases of ytterbium: ferromagnetism and
  mott states in an {SU}(6) fermi system},\ }\href
  {https://doi.org/10.1088/1367-2630/11/10/103033} {\bibfield  {journal}
  {\bibinfo  {journal} {New Journal of Physics}\ }\textbf {\bibinfo {volume}
  {11}},\ \bibinfo {pages} {103033} (\bibinfo {year} {2009})}\BibitemShut
  {NoStop}%
\bibitem [{\citenamefont {Laird}\ \emph {et~al.}(2017)\citenamefont {Laird},
  \citenamefont {Shi}, \citenamefont {Parish},\ and\ \citenamefont
  {Levinsen}}]{PhysRevA.96.032701}%
  \BibitemOpen
  \bibfield  {author} {\bibinfo {author} {\bibfnamefont {E.~K.}\ \bibnamefont
  {Laird}}, \bibinfo {author} {\bibfnamefont {Z.-Y.}\ \bibnamefont {Shi}},
  \bibinfo {author} {\bibfnamefont {M.~M.}\ \bibnamefont {Parish}},\ and\
  \bibinfo {author} {\bibfnamefont {J.}~\bibnamefont {Levinsen}},\ }\bibfield
  {title} {\bibinfo {title} {Su($n$) fermions in a one-dimensional harmonic
  trap},\ }\href {https://doi.org/10.1103/PhysRevA.96.032701} {\bibfield
  {journal} {\bibinfo  {journal} {Phys. Rev. A}\ }\textbf {\bibinfo {volume}
  {96}},\ \bibinfo {pages} {032701} (\bibinfo {year} {2017})}\BibitemShut
  {NoStop}%
\bibitem [{\citenamefont {Bistritzer}\ and\ \citenamefont
  {MacDonald}(2011)}]{Bistritzer12233}%
  \BibitemOpen
  \bibfield  {author} {\bibinfo {author} {\bibfnamefont {R.}~\bibnamefont
  {Bistritzer}}\ and\ \bibinfo {author} {\bibfnamefont {A.~H.}\ \bibnamefont
  {MacDonald}},\ }\bibfield  {title} {\bibinfo {title} {Moiré bands in twisted
  double-layer graphene},\ }\href {https://doi.org/10.1073/pnas.1108174108}
  {\bibfield  {journal} {\bibinfo  {journal} {Proceedings of the National
  Academy of Sciences}\ }\textbf {\bibinfo {volume} {108}},\ \bibinfo {pages}
  {12233–12237} (\bibinfo {year} {2011})}\BibitemShut {NoStop}%
\bibitem [{\citenamefont {Cao}\ \emph {et~al.}(2018)\citenamefont {Cao},
  \citenamefont {Fatemi}, \citenamefont {Demir}, \citenamefont {Fang},
  \citenamefont {Tomarken}, \citenamefont {Luo}, \citenamefont
  {Sanchez-Yamagishi}, \citenamefont {Watanabe}, \citenamefont {Taniguchi},
  \citenamefont {Kaxiras}, \citenamefont {Ashoori},\ and\ \citenamefont
  {Jarillo-Herrero}}]{Cao2018}%
  \BibitemOpen
  \bibfield  {author} {\bibinfo {author} {\bibfnamefont {Y.}~\bibnamefont
  {Cao}}, \bibinfo {author} {\bibfnamefont {V.}~\bibnamefont {Fatemi}},
  \bibinfo {author} {\bibfnamefont {A.}~\bibnamefont {Demir}}, \bibinfo
  {author} {\bibfnamefont {S.}~\bibnamefont {Fang}}, \bibinfo {author}
  {\bibfnamefont {S.~L.}\ \bibnamefont {Tomarken}}, \bibinfo {author}
  {\bibfnamefont {J.~Y.}\ \bibnamefont {Luo}}, \bibinfo {author} {\bibfnamefont
  {J.~D.}\ \bibnamefont {Sanchez-Yamagishi}}, \bibinfo {author} {\bibfnamefont
  {K.}~\bibnamefont {Watanabe}}, \bibinfo {author} {\bibfnamefont
  {T.}~\bibnamefont {Taniguchi}}, \bibinfo {author} {\bibfnamefont
  {E.}~\bibnamefont {Kaxiras}}, \bibinfo {author} {\bibfnamefont {R.~C.}\
  \bibnamefont {Ashoori}},\ and\ \bibinfo {author} {\bibfnamefont
  {P.}~\bibnamefont {Jarillo-Herrero}},\ }\bibfield  {title} {\bibinfo {title}
  {Correlated insulator behaviour at half-filling in magic-angle graphene
  superlattices},\ }\href {https://doi.org/10.1038/nature26154} {\bibfield
  {journal} {\bibinfo  {journal} {Nature}\ }\textbf {\bibinfo {volume} {556}},\
  \bibinfo {pages} {80–84} (\bibinfo {year} {2018})}\BibitemShut {NoStop}%
\bibitem [{\citenamefont {Calixto}\ \emph {et~al.}(2025)\citenamefont
  {Calixto}, \citenamefont {Mayorgas},\ and\ \citenamefont
  {Castaños}}]{TBGPhysicaE}%
  \BibitemOpen
  \bibfield  {author} {\bibinfo {author} {\bibfnamefont {M.}~\bibnamefont
  {Calixto}}, \bibinfo {author} {\bibfnamefont {A.}~\bibnamefont {Mayorgas}},\
  and\ \bibinfo {author} {\bibfnamefont {O.}~\bibnamefont {Castaños}},\
  }\bibfield  {title} {\bibinfo {title} {Capturing magic angles in twisted
  bilayer graphene from information theory markers},\ }\href
  {https://doi.org/https://doi.org/10.1016/j.physe.2025.116199} {\bibfield
  {journal} {\bibinfo  {journal} {Physica E: Low-dimensional Systems and
  Nanostructures}\ }\textbf {\bibinfo {volume} {169}},\ \bibinfo {pages}
  {116199} (\bibinfo {year} {2025})}\BibitemShut {NoStop}%
\bibitem [{\citenamefont {Ezawa}\ \emph {et~al.}(2005)\citenamefont {Ezawa},
  \citenamefont {Eliashvili},\ and\ \citenamefont {Tsitsishvili}}]{HamEzawa}%
  \BibitemOpen
  \bibfield  {author} {\bibinfo {author} {\bibfnamefont {Z.~F.}\ \bibnamefont
  {Ezawa}}, \bibinfo {author} {\bibfnamefont {M.}~\bibnamefont {Eliashvili}},\
  and\ \bibinfo {author} {\bibfnamefont {G.}~\bibnamefont {Tsitsishvili}},\
  }\bibfield  {title} {\bibinfo {title} {Ground-state structure in
  $\ensuremath{\nu}=2$ bilayer quantum hall systems},\ }\href
  {https://doi.org/10.1103/PhysRevB.71.125318} {\bibfield  {journal} {\bibinfo
  {journal} {Phys. Rev. B}\ }\textbf {\bibinfo {volume} {71}},\ \bibinfo
  {pages} {125318} (\bibinfo {year} {2005})}\BibitemShut {NoStop}%
\bibitem [{\citenamefont {Schliemann}\ and\ \citenamefont
  {MacDonald}(2000)}]{Schliemann}%
  \BibitemOpen
  \bibfield  {author} {\bibinfo {author} {\bibfnamefont {J.}~\bibnamefont
  {Schliemann}}\ and\ \bibinfo {author} {\bibfnamefont {A.~H.}\ \bibnamefont
  {MacDonald}},\ }\bibfield  {title} {\bibinfo {title} {Bilayer quantum hall
  systems at filling factor
  $\ensuremath{\nu}\phantom{\rule{0ex}{0ex}}=\phantom{\rule{0ex}{0ex}}2$: An
  exact diagonalization study},\ }\href
  {https://doi.org/10.1103/PhysRevLett.84.4437} {\bibfield  {journal} {\bibinfo
   {journal} {Phys. Rev. Lett.}\ }\textbf {\bibinfo {volume} {84}},\ \bibinfo
  {pages} {4437–4440} (\bibinfo {year} {2000})}\BibitemShut {NoStop}%
\bibitem [{\citenamefont {Calixto}\ \emph {et~al.}(2017)\citenamefont
  {Calixto}, \citenamefont {Peón-Nieto},\ and\ \citenamefont
  {Pérez-Romero}}]{PhysRevB.95.235302}%
  \BibitemOpen
  \bibfield  {author} {\bibinfo {author} {\bibfnamefont {M.}~\bibnamefont
  {Calixto}}, \bibinfo {author} {\bibfnamefont {C.}~\bibnamefont
  {Peón-Nieto}},\ and\ \bibinfo {author} {\bibfnamefont {E.}~\bibnamefont
  {Pérez-Romero}},\ }\bibfield  {title} {\bibinfo {title} {Hilbert space and
  ground-state structure of bilayer quantum hall systems at
  $\ensuremath{\nu}=2$},\ }\href {https://doi.org/10.1103/PhysRevB.95.235302}
  {\bibfield  {journal} {\bibinfo  {journal} {Phys. Rev. B}\ }\textbf {\bibinfo
  {volume} {95}},\ \bibinfo {pages} {235302} (\bibinfo {year}
  {2017})}\BibitemShut {NoStop}%
\bibitem [{\citenamefont {Seki}\ and\ \citenamefont
  {Mochizuki}(2016)}]{Seki-Mochizuki_Skyrmion_2016}%
  \BibitemOpen
  \bibfield  {author} {\bibinfo {author} {\bibfnamefont {S.}~\bibnamefont
  {Seki}}\ and\ \bibinfo {author} {\bibfnamefont {M.}~\bibnamefont
  {Mochizuki}},\ }\href {https://doi.org/10.1007/978-3-319-24651-2} {\emph
  {\bibinfo {title} {Skyrmions in Magnetic Materials}}}\ (\bibinfo  {publisher}
  {Springer, Cham},\ \bibinfo {year} {2016})\BibitemShut {NoStop}%
\bibitem [{\citenamefont {Han}(2017)}]{JungHoonHan_Skyrmion_2017}%
  \BibitemOpen
  \bibfield  {author} {\bibinfo {author} {\bibfnamefont {J.~H.}\ \bibnamefont
  {Han}},\ }\href {https://doi.org/10.1007/978-3-319-69246-3} {\emph {\bibinfo
  {title} {Skyrmions in Condensed Matter}}}\ (\bibinfo  {publisher} {Springer,
  Cham},\ \bibinfo {year} {2017})\BibitemShut {NoStop}%
\bibitem [{\citenamefont {Finocchio}\ and\ \citenamefont
  {Panagopoulos}(2021)}]{Finocchio-Panagopoulos_Skyrmion_2021}%
  \BibitemOpen
  \bibinfo {editor} {\bibfnamefont {G.}~\bibnamefont {Finocchio}}\ and\
  \bibinfo {editor} {\bibfnamefont {C.}~\bibnamefont {Panagopoulos}},\ eds.,\
  \href {https://doi.org/10.1016/C2019-0-02206-6} {}\ (\bibinfo  {publisher}
  {Woodhead Publishing},\ \bibinfo {year} {2021})\BibitemShut {NoStop}%
\bibitem [{\citenamefont {Zhang}(2018)}]{Zang_Thesis_Skyrmion_2018}%
  \BibitemOpen
  \bibfield  {author} {\bibinfo {author} {\bibfnamefont {S.}~\bibnamefont
  {Zhang}},\ }\href {https://doi.org/10.1007/978-3-319-98252-6} {\emph
  {\bibinfo {title} {Chiral and Topological Nature of Magnetic Skyrmions}}}\
  (\bibinfo  {publisher} {Springer, Cham},\ \bibinfo {year} {2018})\BibitemShut
  {NoStop}%
\bibitem [{\citenamefont {Affleck}(1985)}]{AffleckNPB257}%
  \BibitemOpen
  \bibfield  {author} {\bibinfo {author} {\bibfnamefont {I.}~\bibnamefont
  {Affleck}},\ }\bibfield  {title} {\bibinfo {title} {The quantum hall effects,
  $\sigma$-models at $\theta=\pi$ and quantum spin chains},\ }\href
  {https://doi.org/10.1016/0550-3213(85)90353-0} {\bibfield  {journal}
  {\bibinfo  {journal} {Nuclear Physics B}\ }\textbf {\bibinfo {volume}
  {257}},\ \bibinfo {pages} {397–406} (\bibinfo {year} {1985})}\BibitemShut
  {NoStop}%
\bibitem [{\citenamefont {Affleck}(1986)}]{AffleckNPB265}%
  \BibitemOpen
  \bibfield  {author} {\bibinfo {author} {\bibfnamefont {I.}~\bibnamefont
  {Affleck}},\ }\bibfield  {title} {\bibinfo {title} {Exact critical exponents
  for quantum spin chains, non-linear $\sigma$-models at $\theta=\pi$ and the
  quantum hall effect},\ }\href {https://doi.org/10.1016/0550-3213(86)90167-7}
  {\bibfield  {journal} {\bibinfo  {journal} {Nuclear Physics B}\ }\textbf
  {\bibinfo {volume} {265}},\ \bibinfo {pages} {409–447} (\bibinfo {year}
  {1986})}\BibitemShut {NoStop}%
\bibitem [{\citenamefont {Affleck}(1988)}]{AffleckNPB305}%
  \BibitemOpen
  \bibfield  {author} {\bibinfo {author} {\bibfnamefont {I.}~\bibnamefont
  {Affleck}},\ }\bibfield  {title} {\bibinfo {title} {Critical behaviour of
  su$(n)$ quantum chains and topological non-linear $\sigma$-models},\ }\href
  {https://doi.org/10.1016/0550-3213(88)90117-4} {\bibfield  {journal}
  {\bibinfo  {journal} {Nuclear Physics B}\ }\textbf {\bibinfo {volume}
  {305}},\ \bibinfo {pages} {582–596} (\bibinfo {year} {1988})}\BibitemShut
  {NoStop}%
\bibitem [{\citenamefont {Read}\ and\ \citenamefont {Sachdev}(1989)}]{Sachdev}%
  \BibitemOpen
  \bibfield  {author} {\bibinfo {author} {\bibfnamefont {N.}~\bibnamefont
  {Read}}\ and\ \bibinfo {author} {\bibfnamefont {S.}~\bibnamefont {Sachdev}},\
  }\bibfield  {title} {\bibinfo {title} {Some features of the phase diagram of
  the square lattice su(n) antiferromagnet},\ }\href
  {https://doi.org/10.1016/0550-3213(89)90061-8} {\bibfield  {journal}
  {\bibinfo  {journal} {Nuclear Physics B}\ }\textbf {\bibinfo {volume}
  {316}},\ \bibinfo {pages} {609–640} (\bibinfo {year} {1989})}\BibitemShut
  {NoStop}%
\bibitem [{\citenamefont {Read}\ and\ \citenamefont
  {Sachdev}(1990)}]{Sachdev2}%
  \BibitemOpen
  \bibfield  {author} {\bibinfo {author} {\bibfnamefont {N.}~\bibnamefont
  {Read}}\ and\ \bibinfo {author} {\bibfnamefont {S.}~\bibnamefont {Sachdev}},\
  }\bibfield  {title} {\bibinfo {title} {Spin-peierls, valence-bond solid, and
  néel ground states of low-dimensional quantum antiferromagnets},\ }\href
  {https://doi.org/10.1103/PhysRevB.42.4568} {\bibfield  {journal} {\bibinfo
  {journal} {Phys. Rev. B}\ }\textbf {\bibinfo {volume} {42}},\ \bibinfo
  {pages} {4568–4589} (\bibinfo {year} {1990})}\BibitemShut {NoStop}%
\bibitem [{\citenamefont {Arovas}\ \emph {et~al.}(1999)\citenamefont {Arovas},
  \citenamefont {Karlhede},\ and\ \citenamefont {Lilliehöök}}]{Arovas}%
  \BibitemOpen
  \bibfield  {author} {\bibinfo {author} {\bibfnamefont {D.~P.}\ \bibnamefont
  {Arovas}}, \bibinfo {author} {\bibfnamefont {A.}~\bibnamefont {Karlhede}},\
  and\ \bibinfo {author} {\bibfnamefont {D.}~\bibnamefont {Lilliehöök}},\
  }\bibfield  {title} {\bibinfo {title} {$\mathrm{SU}(n)$ quantum hall
  skyrmions},\ }\href {https://doi.org/10.1103/PhysRevB.59.13147} {\bibfield
  {journal} {\bibinfo  {journal} {Phys. Rev. B}\ }\textbf {\bibinfo {volume}
  {59}},\ \bibinfo {pages} {13147–13150} (\bibinfo {year}
  {1999})}\BibitemShut {NoStop}%
\bibitem [{\citenamefont {Calixto}\ \emph {et~al.}(2016)\citenamefont
  {Calixto}, \citenamefont {Peón-Nieto},\ and\ \citenamefont
  {Pérez-Romero}}]{CALIXTO1}%
  \BibitemOpen
  \bibfield  {author} {\bibinfo {author} {\bibfnamefont {M.}~\bibnamefont
  {Calixto}}, \bibinfo {author} {\bibfnamefont {C.}~\bibnamefont
  {Peón-Nieto}},\ and\ \bibinfo {author} {\bibfnamefont {E.}~\bibnamefont
  {Pérez-Romero}},\ }\bibfield  {title} {\bibinfo {title} {Coherent states for
  n-component fractional quantum hall systems and their nonlinear sigma
  models},\ }\href {https://doi.org/10.1016/j.aop.2016.06.025} {\bibfield
  {journal} {\bibinfo  {journal} {Annals of Physics}\ }\textbf {\bibinfo
  {volume} {373}},\ \bibinfo {pages} {52–66} (\bibinfo {year}
  {2016})}\BibitemShut {NoStop}%
\bibitem [{\citenamefont {Haldane}(1983{\natexlab{a}})}]{HaldanePLA93}%
  \BibitemOpen
  \bibfield  {author} {\bibinfo {author} {\bibfnamefont {F.}~\bibnamefont
  {Haldane}},\ }\bibfield  {title} {\bibinfo {title} {Continuum dynamics of the
  1-d heisenberg antiferromagnet: Identification with the o(3) nonlinear sigma
  model},\ }\href {https://doi.org/10.1016/0375-9601(83)90631-X} {\bibfield
  {journal} {\bibinfo  {journal} {Physics Letters A}\ }\textbf {\bibinfo
  {volume} {93}},\ \bibinfo {pages} {464–468} (\bibinfo {year}
  {1983}{\natexlab{a}})}\BibitemShut {NoStop}%
\bibitem [{\citenamefont {Haldane}(1983{\natexlab{b}})}]{HaldanePLA93-2}%
  \BibitemOpen
  \bibfield  {author} {\bibinfo {author} {\bibfnamefont {F.~D.~M.}\
  \bibnamefont {Haldane}},\ }\bibfield  {title} {\bibinfo {title} {Nonlinear
  field theory of large-spin heisenberg antiferromagnets: Semiclassically
  quantized solitons of the one-dimensional easy-axis néel state},\ }\href
  {https://doi.org/10.1103/PhysRevLett.50.1153} {\bibfield  {journal} {\bibinfo
   {journal} {Phys. Rev. Lett.}\ }\textbf {\bibinfo {volume} {50}},\ \bibinfo
  {pages} {1153–1156} (\bibinfo {year} {1983}{\natexlab{b}})}\BibitemShut
  {NoStop}%
\bibitem [{\citenamefont {Haldane}(1988)}]{HaldanePLA93-3}%
  \BibitemOpen
  \bibfield  {author} {\bibinfo {author} {\bibfnamefont {F.~D.~M.}\
  \bibnamefont {Haldane}},\ }\bibfield  {title} {\bibinfo {title} {O(3)
  nonlinear $\ensuremath{\sigma}$ model and the topological distinction between
  integer- and half-integer-spin antiferromagnets in two dimensions},\ }\href
  {https://doi.org/10.1103/PhysRevLett.61.1029} {\bibfield  {journal} {\bibinfo
   {journal} {Phys. Rev. Lett.}\ }\textbf {\bibinfo {volume} {61}},\ \bibinfo
  {pages} {1029–1032} (\bibinfo {year} {1988})}\BibitemShut {NoStop}%
\bibitem [{\citenamefont {Perelomov}(1986)}]{Perelomov}%
  \BibitemOpen
  \bibfield  {author} {\bibinfo {author} {\bibfnamefont {A.}~\bibnamefont
  {Perelomov}},\ }\href {https://doi.org/10.1007/978-3-642-61629-7} {\emph
  {\bibinfo {title} {Generalized Coherent States and their Applications}}}\
  (\bibinfo  {publisher} {Springer-Verlag Berlin Heidelberg},\ \bibinfo {year}
  {1986})\BibitemShut {NoStop}%
\bibitem [{\citenamefont {Gazeau}(2009)}]{Gazeaubook}%
  \BibitemOpen
  \bibfield  {author} {\bibinfo {author} {\bibfnamefont {J.}~\bibnamefont
  {Gazeau}},\ }\href {https://doi.org/10.1002/9783527628285} {}\ (\bibinfo
  {publisher} {John Wiley \& Sons, Ltd},\ \bibinfo {year} {2009})\BibitemShut
  {NoStop}%
\bibitem [{\citenamefont {Calixto}\ \emph
  {et~al.}(2021{\natexlab{a}})\citenamefont {Calixto}, \citenamefont
  {Mayorgas},\ and\ \citenamefont {Guerrero}}]{QIP-2021-Entanglement}%
  \BibitemOpen
  \bibfield  {author} {\bibinfo {author} {\bibfnamefont {M.}~\bibnamefont
  {Calixto}}, \bibinfo {author} {\bibfnamefont {A.}~\bibnamefont {Mayorgas}},\
  and\ \bibinfo {author} {\bibfnamefont {J.}~\bibnamefont {Guerrero}},\
  }\bibfield  {title} {\bibinfo {title} {Entanglement and {U(D)}-spin squeezing
  in symmetric multi-qudit systems and applications to quantum phase
  transitions in {L}ipkin–{M}eshkov–{G}lick {D}-level atom models},\ }\href
  {https://doi.org/10.1007/s11128-021-03218-6} {\bibfield  {journal} {\bibinfo
  {journal} {Quantum Information Processing}\ }\textbf {\bibinfo {volume}
  {20}},\ \bibinfo {pages} {304} (\bibinfo {year}
  {2021}{\natexlab{a}})}\BibitemShut {NoStop}%
\bibitem [{\citenamefont {Frame}\ \emph {et~al.}(1954)\citenamefont {Frame},
  \citenamefont {Robinson},\ and\ \citenamefont {Thrall}}]{Thrall1954}%
  \BibitemOpen
  \bibfield  {author} {\bibinfo {author} {\bibfnamefont {J.~S.}\ \bibnamefont
  {Frame}}, \bibinfo {author} {\bibfnamefont {G.~d.~B.}\ \bibnamefont
  {Robinson}},\ and\ \bibinfo {author} {\bibfnamefont {R.~M.}\ \bibnamefont
  {Thrall}},\ }\bibfield  {title} {\bibinfo {title} {The hook graphs of the
  symmetric group},\ }\href {https://doi.org/10.4153/CJM-1954-030-1} {\bibfield
   {journal} {\bibinfo  {journal} {Canadian Journal of Mathematics}\ }\textbf
  {\bibinfo {volume} {6}},\ \bibinfo {pages} {316–324} (\bibinfo {year}
  {1954})}\BibitemShut {NoStop}%
\bibitem [{\citenamefont {Barut}\ and\ \citenamefont {Raczka}(1980)}]{Barut}%
  \BibitemOpen
  \bibfield  {author} {\bibinfo {author} {\bibfnamefont {A.}~\bibnamefont
  {Barut}}\ and\ \bibinfo {author} {\bibfnamefont {R.}~\bibnamefont {Raczka}},\
  }\href {https://www.worldscientific.com/doi/10.1142/0352} {\emph {\bibinfo
  {title} {Theory of Group Representations and Applications}}}\ (\bibinfo
  {publisher} {Polish Scientific Publishers, Warszawa},\ \bibinfo {year}
  {1980})\BibitemShut {NoStop}%
\bibitem [{\citenamefont {Hall}(2004)}]{HallBook}%
  \BibitemOpen
  \bibfield  {author} {\bibinfo {author} {\bibfnamefont {B.}~\bibnamefont
  {Hall}},\ }\href {https://link.springer.com/book/10.1007/978-3-319-13467-3}
  {\emph {\bibinfo {title} {Lie Groups, Lie Algebras, and Representations}}}\
  (\bibinfo  {publisher} {Springer},\ \bibinfo {year} {2004})\BibitemShut
  {NoStop}%
\bibitem [{\citenamefont {Cvitanovic}(2008)}]{CVI}%
  \BibitemOpen
  \bibfield  {author} {\bibinfo {author} {\bibfnamefont {P.}~\bibnamefont
  {Cvitanovic}},\ }\href
  {https://press.princeton.edu/books/paperback/9780691202983/group-theory}
  {\emph {\bibinfo {title} {Group theory: Birdtracks, Lie's and exceptional
  groups}}}\ (\bibinfo  {publisher} {Princeton University Press},\ \bibinfo
  {year} {2008})\BibitemShut {NoStop}%
\bibitem [{\citenamefont {Zhao}(2008)}]{ZHAO}%
  \BibitemOpen
  \bibfield  {author} {\bibinfo {author} {\bibfnamefont {Y.}~\bibnamefont
  {Zhao}},\ }\href {https://yufeizhao.com/research/youngtab-hcmr.pdf} {\bibinfo
  {title} {Young tableaux and the representations of the symmetric group}},\
  \bibinfo {howpublished} {Student article, Massachusetts Institute of
  Technology, USA} (\bibinfo {year} {2008})\BibitemShut {NoStop}%
\bibitem [{\citenamefont {Zhang}\ \emph {et~al.}(1990)\citenamefont {Zhang},
  \citenamefont {Feng},\ and\ \citenamefont {Gilmore}}]{RevModPhys.62.867}%
  \BibitemOpen
  \bibfield  {author} {\bibinfo {author} {\bibfnamefont {W.-M.}\ \bibnamefont
  {Zhang}}, \bibinfo {author} {\bibfnamefont {D.~H.}\ \bibnamefont {Feng}},\
  and\ \bibinfo {author} {\bibfnamefont {R.}~\bibnamefont {Gilmore}},\
  }\bibfield  {title} {\bibinfo {title} {Coherent states: Theory and some
  applications},\ }\href {https://doi.org/10.1103/RevModPhys.62.867} {\bibfield
   {journal} {\bibinfo  {journal} {Rev. Mod. Phys.}\ }\textbf {\bibinfo
  {volume} {62}},\ \bibinfo {pages} {867–927} (\bibinfo {year}
  {1990})}\BibitemShut {NoStop}%
\bibitem [{\citenamefont {Glauber}(1963)}]{PhysRev.131.2766}%
  \BibitemOpen
  \bibfield  {author} {\bibinfo {author} {\bibfnamefont {R.~J.}\ \bibnamefont
  {Glauber}},\ }\bibfield  {title} {\bibinfo {title} {Coherent and incoherent
  states of the radiation field},\ }\href
  {https://doi.org/10.1103/PhysRev.131.2766} {\bibfield  {journal} {\bibinfo
  {journal} {Phys. Rev.}\ }\textbf {\bibinfo {volume} {131}},\ \bibinfo {pages}
  {2766–2788} (\bibinfo {year} {1963})}\BibitemShut {NoStop}%
\bibitem [{\citenamefont {Arecchi}\ \emph
  {et~al.}(1972{\natexlab{a}})\citenamefont {Arecchi}, \citenamefont
  {Courtens}, \citenamefont {Gilmore},\ and\ \citenamefont
  {Thomas}}]{GilmorePhysRevA.6.2211}%
  \BibitemOpen
  \bibfield  {author} {\bibinfo {author} {\bibfnamefont {F.~T.}\ \bibnamefont
  {Arecchi}}, \bibinfo {author} {\bibfnamefont {E.}~\bibnamefont {Courtens}},
  \bibinfo {author} {\bibfnamefont {R.}~\bibnamefont {Gilmore}},\ and\ \bibinfo
  {author} {\bibfnamefont {H.}~\bibnamefont {Thomas}},\ }\bibfield  {title}
  {\bibinfo {title} {Atomic coherent states in quantum optics},\ }\href
  {https://doi.org/10.1103/PhysRevA.6.2211} {\bibfield  {journal} {\bibinfo
  {journal} {Phys. Rev. A}\ }\textbf {\bibinfo {volume} {6}},\ \bibinfo {pages}
  {2211–2237} (\bibinfo {year} {1972}{\natexlab{a}})}\BibitemShut {NoStop}%
\bibitem [{\citenamefont {Calixto}\ \emph
  {et~al.}(2021{\natexlab{b}})\citenamefont {Calixto}, \citenamefont
  {Mayorgas},\ and\ \citenamefont {Guerrero}}]{nuestroPRE}%
  \BibitemOpen
  \bibfield  {author} {\bibinfo {author} {\bibfnamefont {M.}~\bibnamefont
  {Calixto}}, \bibinfo {author} {\bibfnamefont {A.}~\bibnamefont {Mayorgas}},\
  and\ \bibinfo {author} {\bibfnamefont {J.}~\bibnamefont {Guerrero}},\
  }\bibfield  {title} {\bibinfo {title} {Role of mixed permutation symmetry
  sectors in the thermodynamic limit of critical three-level
  {Lipkin-Meshkov-Glick} atom models},\ }\href
  {https://doi.org/10.1103/PhysRevE.103.012116} {\bibfield  {journal} {\bibinfo
   {journal} {Phys. Rev. E}\ }\textbf {\bibinfo {volume} {103}},\ \bibinfo
  {pages} {012116} (\bibinfo {year} {2021}{\natexlab{b}})}\BibitemShut
  {NoStop}%
\bibitem [{\citenamefont {Gilmore}(1981)}]{Gilmorebook}%
  \BibitemOpen
  \bibfield  {author} {\bibinfo {author} {\bibfnamefont {R.}~\bibnamefont
  {Gilmore}},\ }\href
  {https://onlinelibrary.wiley.com/doi/abs/10.1002/3527600434.eap052.pub2}
  {\emph {\bibinfo {title} {Catastrophe Theory for Scientists and Engineers}}}\
  (\bibinfo  {publisher} {Wiley, New York},\ \bibinfo {year}
  {1981})\BibitemShut {NoStop}%
\bibitem [{\citenamefont {Lieb}\ and\ \citenamefont
  {Mattis}(1962)}]{Lieb-Mattis_PR1962}%
  \BibitemOpen
  \bibfield  {author} {\bibinfo {author} {\bibfnamefont {E.}~\bibnamefont
  {Lieb}}\ and\ \bibinfo {author} {\bibfnamefont {D.}~\bibnamefont {Mattis}},\
  }\bibfield  {title} {\bibinfo {title} {Theory of ferromagnetism and the
  ordering of electronic energy levels},\ }\href
  {https://doi.org/10.1103/PhysRev.125.164} {\bibfield  {journal} {\bibinfo
  {journal} {Phys. Rev.}\ }\textbf {\bibinfo {volume} {125}},\ \bibinfo {pages}
  {164–172} (\bibinfo {year} {1962})}\BibitemShut {NoStop}%
\bibitem [{\citenamefont {Decamp}\ \emph
  {et~al.}(2020{\natexlab{b}})\citenamefont {Decamp}, \citenamefont {Gong},
  \citenamefont {Loh},\ and\ \citenamefont {Miniatura}}]{Decamp_PRR2020}%
  \BibitemOpen
  \bibfield  {author} {\bibinfo {author} {\bibfnamefont {J.}~\bibnamefont
  {Decamp}}, \bibinfo {author} {\bibfnamefont {J.}~\bibnamefont {Gong}},
  \bibinfo {author} {\bibfnamefont {H.}~\bibnamefont {Loh}},\ and\ \bibinfo
  {author} {\bibfnamefont {C.}~\bibnamefont {Miniatura}},\ }\bibfield  {title}
  {\bibinfo {title} {Graph-theory treatment of one-dimensional strongly
  repulsive fermions},\ }\href
  {https://doi.org/10.1103/PhysRevResearch.2.023059} {\bibfield  {journal}
  {\bibinfo  {journal} {Phys. Rev. Research}\ }\textbf {\bibinfo {volume}
  {2}},\ \bibinfo {pages} {023059} (\bibinfo {year}
  {2020}{\natexlab{b}})}\BibitemShut {NoStop}%
\bibitem [{\citenamefont {Calixto}\ \emph {et~al.}(2022)\citenamefont
  {Calixto}, \citenamefont {Mayorgas},\ and\ \citenamefont
  {Guerrero}}]{sym14050872}%
  \BibitemOpen
  \bibfield  {author} {\bibinfo {author} {\bibfnamefont {M.}~\bibnamefont
  {Calixto}}, \bibinfo {author} {\bibfnamefont {A.}~\bibnamefont {Mayorgas}},\
  and\ \bibinfo {author} {\bibfnamefont {J.}~\bibnamefont {Guerrero}},\
  }\bibfield  {title} {\bibinfo {title} {Hilbert space structure of the low
  energy sector of {U(N)} quantum hall ferromagnets and their classical
  limit},\ }\bibfield  {journal} {\bibinfo  {journal} {Symmetry}\ }\textbf
  {\bibinfo {volume} {14}},\ \href {https://doi.org/10.3390/sym14050872}
  {10.3390/sym14050872} (\bibinfo {year} {2022})\BibitemShut {NoStop}%
\bibitem [{\citenamefont {Casta\~nos}\ \emph {et~al.}(2006)\citenamefont
  {Casta\~nos}, \citenamefont {López-Pe\~na}, \citenamefont {Hirsch},\ and\
  \citenamefont {López-Moreno}}]{octavio}%
  \BibitemOpen
  \bibfield  {author} {\bibinfo {author} {\bibfnamefont {O.}~\bibnamefont
  {Casta\~nos}}, \bibinfo {author} {\bibfnamefont {R.}~\bibnamefont
  {López-Pe\~na}}, \bibinfo {author} {\bibfnamefont {J.~G.}\ \bibnamefont
  {Hirsch}},\ and\ \bibinfo {author} {\bibfnamefont {E.}~\bibnamefont
  {López-Moreno}},\ }\bibfield  {title} {\bibinfo {title} {Classical and
  quantum phase transitions in the {Lipkin-Meshkov-Glick} model},\ }\href
  {https://doi.org/10.1103/PhysRevB.74.104118} {\bibfield  {journal} {\bibinfo
  {journal} {Phys. Rev. B}\ }\textbf {\bibinfo {volume} {74}},\ \bibinfo
  {pages} {104118} (\bibinfo {year} {2006})}\BibitemShut {NoStop}%
\bibitem [{\citenamefont {Romera}\ \emph {et~al.}(2014)\citenamefont {Romera},
  \citenamefont {Calixto},\ and\ \citenamefont {Castaños}}]{Romera_2014}%
  \BibitemOpen
  \bibfield  {author} {\bibinfo {author} {\bibfnamefont {E.}~\bibnamefont
  {Romera}}, \bibinfo {author} {\bibfnamefont {M.}~\bibnamefont {Calixto}},\
  and\ \bibinfo {author} {\bibfnamefont {O.}~\bibnamefont {Castaños}},\
  }\bibfield  {title} {\bibinfo {title} {Phase space analysis of first-,
  second- and third-order quantum phase transitions in the
  {Lipkin-Meshkov-Glick} model},\ }\href
  {https://doi.org/10.1088/0031-8949/89/9/095103} {\bibfield  {journal}
  {\bibinfo  {journal} {Physica Scripta}\ }\textbf {\bibinfo {volume} {89}},\
  \bibinfo {pages} {095103} (\bibinfo {year} {2014})}\BibitemShut {NoStop}%
\bibitem [{\citenamefont {Cordero}\ \emph {et~al.}(2015)\citenamefont
  {Cordero}, \citenamefont {Nahmad-Achar}, \citenamefont {López-Pe\~na},\ and\
  \citenamefont {Casta\~nos}}]{PhysRevA.92.053843}%
  \BibitemOpen
  \bibfield  {author} {\bibinfo {author} {\bibfnamefont {S.}~\bibnamefont
  {Cordero}}, \bibinfo {author} {\bibfnamefont {E.}~\bibnamefont
  {Nahmad-Achar}}, \bibinfo {author} {\bibfnamefont {R.}~\bibnamefont
  {López-Pe\~na}},\ and\ \bibinfo {author} {\bibfnamefont {O.}~\bibnamefont
  {Casta\~nos}},\ }\bibfield  {title} {\bibinfo {title} {Polychromatic phase
  diagram for $n$-level atoms interacting with $l$ modes of an electromagnetic
  field},\ }\href {https://doi.org/10.1103/PhysRevA.92.053843} {\bibfield
  {journal} {\bibinfo  {journal} {Phys. Rev. A}\ }\textbf {\bibinfo {volume}
  {92}},\ \bibinfo {pages} {053843} (\bibinfo {year} {2015})}\BibitemShut
  {NoStop}%
\bibitem [{\citenamefont {Cordero}\ \emph {et~al.}(2016)\citenamefont
  {Cordero}, \citenamefont {Casta\~nos}, \citenamefont {López-Pe\~na},\ and\
  \citenamefont {Nahmad-Achar}}]{PhysRevA.94.013802}%
  \BibitemOpen
  \bibfield  {author} {\bibinfo {author} {\bibfnamefont {S.}~\bibnamefont
  {Cordero}}, \bibinfo {author} {\bibfnamefont {O.}~\bibnamefont {Casta\~nos}},
  \bibinfo {author} {\bibfnamefont {R.}~\bibnamefont {López-Pe\~na}},\ and\
  \bibinfo {author} {\bibfnamefont {E.}~\bibnamefont {Nahmad-Achar}},\
  }\bibfield  {title} {\bibinfo {title} {Variational study of
  $\ensuremath{\lambda}$ and $n$ atomic configurations interacting with an
  electromagnetic field of two modes},\ }\href
  {https://doi.org/10.1103/PhysRevA.94.013802} {\bibfield  {journal} {\bibinfo
  {journal} {Phys. Rev. A}\ }\textbf {\bibinfo {volume} {94}},\ \bibinfo
  {pages} {013802} (\bibinfo {year} {2016})}\BibitemShut {NoStop}%
\bibitem [{\citenamefont {Arecchi}\ \emph
  {et~al.}(1972{\natexlab{b}})\citenamefont {Arecchi}, \citenamefont
  {Courtens}, \citenamefont {Gilmore},\ and\ \citenamefont {Thomas}}]{Gilmore}%
  \BibitemOpen
  \bibfield  {author} {\bibinfo {author} {\bibfnamefont {F.~T.}\ \bibnamefont
  {Arecchi}}, \bibinfo {author} {\bibfnamefont {E.}~\bibnamefont {Courtens}},
  \bibinfo {author} {\bibfnamefont {R.}~\bibnamefont {Gilmore}},\ and\ \bibinfo
  {author} {\bibfnamefont {H.}~\bibnamefont {Thomas}},\ }\bibfield  {title}
  {\bibinfo {title} {Atomic coherent states in quantum optics},\ }\href
  {https://doi.org/10.1103/PhysRevA.6.2211} {\bibfield  {journal} {\bibinfo
  {journal} {Phys. Rev. A}\ }\textbf {\bibinfo {volume} {6}},\ \bibinfo {pages}
  {2211–2237} (\bibinfo {year} {1972}{\natexlab{b}})}\BibitemShut {NoStop}%
\bibitem [{\citenamefont {{Lipkin}}\ \emph
  {et~al.}(1965{\natexlab{a}})\citenamefont {{Lipkin}}, \citenamefont
  {{Meshkov}},\ and\ \citenamefont {{Glick}}}]{lipkin1}%
  \BibitemOpen
  \bibfield  {author} {\bibinfo {author} {\bibfnamefont {H.~J.}\ \bibnamefont
  {{Lipkin}}}, \bibinfo {author} {\bibfnamefont {N.}~\bibnamefont
  {{Meshkov}}},\ and\ \bibinfo {author} {\bibfnamefont {A.~J.}\ \bibnamefont
  {{Glick}}},\ }\bibfield  {title} {\bibinfo {title} {Validity of many-body
  approximation methods for a solvable model. (i). exact solutions and
  perturbation theory},\ }\href {https://doi.org/10.1016/0029-5582(65)90862-X}
  {\bibfield  {journal} {\bibinfo  {journal} {Nuclear Physics}\ }\textbf
  {\bibinfo {volume} {62}},\ \bibinfo {pages} {188–198} (\bibinfo {year}
  {1965}{\natexlab{a}})}\BibitemShut {NoStop}%
\bibitem [{\citenamefont {{Lipkin}}\ \emph
  {et~al.}(1965{\natexlab{b}})\citenamefont {{Lipkin}}, \citenamefont
  {{Meshkov}},\ and\ \citenamefont {{Glick}}}]{lipkin2}%
  \BibitemOpen
  \bibfield  {author} {\bibinfo {author} {\bibfnamefont {H.~J.}\ \bibnamefont
  {{Lipkin}}}, \bibinfo {author} {\bibfnamefont {N.}~\bibnamefont
  {{Meshkov}}},\ and\ \bibinfo {author} {\bibfnamefont {A.~J.}\ \bibnamefont
  {{Glick}}},\ }\bibfield  {title} {\bibinfo {title} {Validity of many-body
  approximation methods for a solvable model: (iii). diagram summations},\
  }\href {https://doi.org/10.1016/0029-5582(65)90864-3} {\bibfield  {journal}
  {\bibinfo  {journal} {Nuclear Physics}\ }\textbf {\bibinfo {volume} {62}},\
  \bibinfo {pages} {211–224} (\bibinfo {year}
  {1965}{\natexlab{b}})}\BibitemShut {NoStop}%
\bibitem [{\citenamefont {Caprio}\ \emph {et~al.}(2008)\citenamefont {Caprio},
  \citenamefont {Cejnar},\ and\ \citenamefont {Iachello}}]{ESQPT}%
  \BibitemOpen
  \bibfield  {author} {\bibinfo {author} {\bibfnamefont {M.~A.}\ \bibnamefont
  {Caprio}}, \bibinfo {author} {\bibfnamefont {P.}~\bibnamefont {Cejnar}},\
  and\ \bibinfo {author} {\bibfnamefont {F.}~\bibnamefont {Iachello}},\
  }\bibfield  {title} {\bibinfo {title} {Excited state quantum phase
  transitions in many-body systems},\ }\href
  {http://www.sciencedirect.com/science/article/pii/S0003491607001042}
  {\bibfield  {journal} {\bibinfo  {journal} {Annals of Physics}\ }\textbf
  {\bibinfo {volume} {323}},\ \bibinfo {pages} {1106–1135} (\bibinfo {year}
  {2008})}\BibitemShut {NoStop}%
\bibitem [{\citenamefont {Pérez-Fernández}\ \emph {et~al.}(2011)\citenamefont
  {Pérez-Fernández}, \citenamefont {Rela\~no}, \citenamefont {Arias},
  \citenamefont {Cejnar}, \citenamefont {Dukelsky},\ and\ \citenamefont
  {García-Ramos}}]{relano}%
  \BibitemOpen
  \bibfield  {author} {\bibinfo {author} {\bibfnamefont {P.}~\bibnamefont
  {Pérez-Fernández}}, \bibinfo {author} {\bibfnamefont {A.}~\bibnamefont
  {Rela\~no}}, \bibinfo {author} {\bibfnamefont {J.~M.}\ \bibnamefont {Arias}},
  \bibinfo {author} {\bibfnamefont {P.}~\bibnamefont {Cejnar}}, \bibinfo
  {author} {\bibfnamefont {J.}~\bibnamefont {Dukelsky}},\ and\ \bibinfo
  {author} {\bibfnamefont {J.~E.}\ \bibnamefont {García-Ramos}},\ }\bibfield
  {title} {\bibinfo {title} {Excited-state phase transition and onset of chaos
  in quantum optical models},\ }\href
  {https://doi.org/10.1103/PhysRevE.83.046208} {\bibfield  {journal} {\bibinfo
  {journal} {Phys. Rev. E}\ }\textbf {\bibinfo {volume} {83}},\ \bibinfo
  {pages} {046208} (\bibinfo {year} {2011})}\BibitemShut {NoStop}%
\bibitem [{\citenamefont {Greiner}\ and\ \citenamefont
  {Müller}(1994)}]{GREINER}%
  \BibitemOpen
  \bibfield  {author} {\bibinfo {author} {\bibfnamefont {W.}~\bibnamefont
  {Greiner}}\ and\ \bibinfo {author} {\bibfnamefont {B.}~\bibnamefont
  {Müller}},\ }\href {https://www.springer.com/gp/book/9783642579769} {\emph
  {\bibinfo {title} {Quantum Mechanics. Symmetries}}}\ (\bibinfo  {publisher}
  {Springer},\ \bibinfo {year} {1994})\BibitemShut {NoStop}%
\bibitem [{\citenamefont {Mayorgas}(2025{\natexlab{a}})}]{Supplemental}%
  \BibitemOpen
  \bibfield  {author} {\bibinfo {author} {\bibfnamefont {A.}~\bibnamefont
  {Mayorgas}},\ }\href {https://doi.org/10.5281/zenodo.15426376} {\bibinfo
  {title} {Supplementary material of lieb-mattis ordering theorem of electronic
  energy levels in the thermodynamic limit}} (\bibinfo {year}
  {2025}{\natexlab{a}})\BibitemShut {NoStop}%
\bibitem [{\citenamefont {Mayorgas}(2025{\natexlab{b}})}]{DiagonalizationData}%
  \BibitemOpen
  \bibfield  {author} {\bibinfo {author} {\bibfnamefont {A.}~\bibnamefont
  {Mayorgas}},\ }\href {https://doi.org/10.5281/zenodo.15583367} {\bibinfo
  {title} {Ground state energy as a function of lambda for different unitary
  representations of the 3-level lmg model}} (\bibinfo {year}
  {2025}{\natexlab{b}})\BibitemShut {NoStop}%
\end{thebibliography}%

\end{document}